%% file: main.tex
\documentclass[prd,aps,showpacs,tightenlines,10pt,reprint,nofootinbib,superscriptaddress, floatfix,twocolumn]{revtex4-2}

\input{packages.tex}

\input{commands.tex}

\begin{document}

\title{Fermion Proca Stars: Vector Dark Matter Admixed Neutron Stars}

\author{Cédric Jockel\,\orcidlink{0009-0007-7617-7178}}
\email{cedric.jockel@aei.mpg.de}
\affiliation{Max Planck Institute for Gravitational Physics (Albert Einstein Institute), Am Mühlenberg 1, 14476 Potsdam, Germany, European Union}
\affiliation{Institute for Theoretical Physics, Goethe University, 60438 Frankfurt am Main, Germany}

\author{Laura Sagunski\,\orcidlink{0000-0002-3506-3306}}
\email{sagunski@itp.uni-frankfurt.de}
\affiliation{Institute for Theoretical Physics, Goethe University, 60438 Frankfurt am Main, Germany}

\date{\today}

\begin{abstract}
Dark matter could accumulate around neutron stars in sufficient amounts to affect their global properties. In this work, we study the effect of a specific model for dark matter -- a massive and self-interacting vector (spin-1) field -- on neutron stars. We describe the combined systems of neutron stars and vector dark matter using Einstein-Proca theory coupled to a nuclear-matter term, and find scaling relations between the field and metric components in the equations of motion. We construct equilibrium solutions of the combined systems, compute their masses and radii and also analyse their stability and higher modes. The combined systems admit dark matter (DM) core and cloud solutions. Core solutions compactify the neutron star component and tend to decrease the total mass of the combined system. Cloud solutions have the inverse effect. Electromagnetic observations of certain cloud-like configurations would appear to violate the Buchdahl limit. This could make Buchdahl-limit violating objects smoking gun signals for dark matter in neutron stars. The self-interaction strength is found to significantly affect both mass and radius. We also compare fermion Proca stars to objects where the dark matter is modelled using a complex scalar field. We find that fermion Proca stars tend to be more massive and geometrically larger than their scalar field counterparts for equal boson masses and self-interaction strengths. Both systems can produce degenerate masses and radii for different amounts of DM and DM particle masses.
\end{abstract}

\maketitle

\section{Introduction \label{sec:intro}} 
\input{introduction.tex}
\section{Theoretical Background \label{sec:theory}} 
\input{theory.tex}
\section{Results \label{sec:Results}} 
\input{results.tex}
\section{Conclusions \label{sec:conclusions}}
\input{conclusions.tex}

\begin{acknowledgments}
The authors acknowledge support by the Deutsche Forschungsgemeinschaft (DFG, German Research Foundation) through the CRC-TR 211 `Strong-interaction matter under extreme conditions'– project number 315477589-TRR 211. CJ acknowledges support by the Hermann-Wilkomm-Stiftung 2023.
\end{acknowledgments}
\appendix
\input{appendix.tex}


\bibliographystyle{apsrev4-1}
\bibliography{biblio.bib}{}

\end{document}

%% file: packages.tex
\usepackage[utf8]{inputenc}
\usepackage{amssymb}
\usepackage{tabularx}
\usepackage{multirow}
\usepackage{hyperref}
\usepackage{amsmath}
\usepackage{amsthm}  
\usepackage[left=2.5cm,right=2.5cm,top=2.5cm,bottom=2.5cm]{geometry}
\usepackage{tensor}
\usepackage{mathrsfs}
\usepackage{enumitem}
\usepackage{lipsum}
\usepackage{comment}
\usepackage[dvipsnames]{xcolor} 
\usepackage{graphicx} 
\usepackage{orcidlink} 

\newcommand{\e}[1]{\cdot 10^{#1}} 

%% file: commands.tex
\renewcommand{\eqref}[1]{Eq.~(\ref{#1})}

%% file: introduction.tex
The nature of dark matter (DM) is one of the large remaining open questions in physics. Even though it constitutes roughly $26.8\%$ of the total energy density of the universe \cite{Planck:2015fie} and has a long observational history \cite{Bertone:2016nfn}, its properties remain largely unknown. We currently know that DM likely is a particle that is only interacting gravitationally and weakly with standard model particles, and that is invisible through electromagnetic radiation. Large-scale structure formation in the universe further suggests that DM is mostly cold, i.e., slowly moving \cite{Bertone:2016nfn,White:1983fcs,Blumenthal:1984bp,Davis:1985rj}. This makes it an integral part of the standard model of cosmology.\\

Neutron stars (NSs) are used to probe a large range of physical phenomena. They are dense and compact remnants of heavy stars. Their high densities make them excellent laboratories for probing gravitation and nuclear physics under extreme conditions. They are characterized using the nuclear matter equation of state (EOS). The EOS describes the relation between pressure and energy density of the matter found inside NSs. It is needed to close the system of differential equations -- the Tolman-Oppenheimer-Volkoff (TOV) equations \cite{Tolman:1939jz,Oppenheimer:1939ne} -- that describe the density distribution of a spherically symmetric static NS and the spacetime curvature. A significant constraint on the EOS is the ability to produce NSs with masses larger than two solar masses, $2\,M_\odot$. The most massive NS known to date is PSR J0952\ensuremath{-}0607 with a mass of $M=2.35^{+0.17}_{-0.17}\,M_\odot$ \cite{Romani:2022jhd}. The lighter companion of the binary system observed in the GW190814 gravitational wave event \cite{LIGOScientific:2020zkf} was also proposed to be the heaviest NS, with a mass of around $2.6\,M_\odot$. But there is evidence that it might be the lightest known black hole instead \cite{Most:2020bba}. High maximum NS masses require stiff EOS, where the nuclear matter is difficult to compress and the energy density rises sharply with increasing pressure. Other constraints include the measurements of the pulsars PSR J0030+0451 \cite{Riley:2019yda} and J0740+6620 \cite{Riley:2021pdl} by the NICER telescope. They also favor a stiff EOS. In contrast, the gravitational wave event GW170817 \cite{Abbott:2018exr,Abbott:2018wiz} favors soft EOS which produce smaller NSs that are more compact and more difficult to tidally disrupt. \\

Additionally, it has been proposed to probe the DM properties using NSs. For example, DM can form a cloud or accumulate inside NSs as a core. In sufficient amounts, it can modify the NS properties such as mass, radius and tidal deformability. These properties have been measured using telescopes such as NICER and the gravitational wave detectors LIGO, Virgo and KAGRA. This allows us to probe the properties of DM such as its particle mass and self-interaction strength (see, e.g., \cite{Diedrichs:2023trk,Goldman:1989nd,Rutherford:2022xeb,Bertone:2007ae,Ellis:2018bkr,Shirke:2023ktu,Karkevandi:2021ygv}). There exist numerous candidates for DM particles. A possible DM candidate is an additional bosonic field (scalar field or vector field), as was studied in \cite{Khlopov:1985jw,Ferreira:2020fam,Biswas:2022tcw,An:2023mvf,GRAVITY:2023azi}. \\

The idea that an astrophysical object consists of a mixture of fermionic and bosonic matter goes back to \cite{Henriques:1989ar,Henriques:1989ez}. A multitude of different models of these fermion boson stars (FBSs) have since been investigated (see, e.g., \cite{Kouvaris:2010vv,Liebling:2012fv,Bertone:2007ae,Zurek:2013wia,Kouvaris:2013awa} for reviews). In the simplest case, the fermionic and bosonic components interact only gravitationally through the effects of their matter-energy content and without an explicit coupling (i.e., they are minimally coupled). This makes FBSs interesting objects in the context of DM research (see, e.g., \cite{Diedrichs:2023trk,DiGiovanni:2021ejn,Nelson:2018xtr}). They have been studied in connection to NSs, where the NS provides the fermionic component and a bosonic field provides the bosonic component of the FBS \cite{Diedrichs:2023trk,DiGiovanni:2021ejn}. The bosonic component can be modelled via, e.g., scalar and vector fields. Another possibility is to study the bosonic field as a fluid or gas of particles. Which treatment is used depends mainly on the mass range of the supposed DM particle. With DM masses above the $eV$ scale, it is treated as a collection of particles or as a fluid. For DM masses $\ll eV$ the correlation length becomes larger than the average particle separation. Dark matter is then best described as a macroscopic wave. We follow the second approach in this work. We refer to \cite{Baryakhtar:2022hbu} for a review of observational prospects of DM at different mass scales. \\

FBSs have been studied with regard to their stability \cite{Henriques:1989ez}. Their dynamical properties were explored in \cite{DiGiovanni:2020frc,DiGiovanni:2021vlu,Valdez-Alvarado:2012rct,Valdez-Alvarado:2020vqa,Ruter:2023uzc,Bezares:2019jcb}. Numerical simulations aiming to understand the gravitational wave signals were performed by \cite{Bezares:2019jcb}. In all these cases, the NS component was modelled using a perfect fluid and a classical complex scalar field was used for the bosonic component. However, understanding vector fields is equally relevant for a number of reasons. If DM is a spin-1 particle, it would be described using a vector field. Some theories of modified gravity also feature vector fields with similar behavior \cite{Siemonsen:2022ivj,Silva:2021jya,Ramazanoglu:2017xbl,Kase:2020yhw,Minamitsuji:2020pak}. In these cases, the vector field is usually directly coupled to the curvature. But if the coupling is weak enough, these fields could behave very similar to minimally coupled fields. In this work, we therefore explore the effect of minimally coupled vector fields on NSs. \\

Fermion boson stars can form in a variety of ways. But in essence, the problem reduces to how one can accumulate a large amount of scalar or vector fields in and around a NS. One common motivation for these fields is bosonic dark matter. It could arrange itself around NSs as a cloud or inside NSs as a core. \\
NSs with DM cores could form 
\begin{itemize}
\itemsep0em
\item[$1)$] from an initial DM 'seed' through accretion of baryonic matter \cite{Diedrichs:2023trk,Meliani:2016rfe,Ellis:2017jgp,Kamenetskaia:2022lbf},
\item[$2)$] through mergers of NSs and boson stars \cite{Diedrichs:2023trk},
\item[$3)$] through accretion of DM onto a NS and subsequent accumulation in the center \cite{Diedrichs:2023trk,Brito:2015yga,Kouvaris:2010vv,Goldman:1989nd,Kouvaris:2007ay},
\item[$4)$] through the decay of standard model particles inside the NS into DM \cite{Baym:2018ljz,Motta:2018rxp,Motta:2018bil,Husain:2022bxl,Berryman:2022zic}.
\end{itemize}

Points $3)$ and $4)$ in particular are highly model dependant. Accretion and decay rates depend on the DM particle interaction cross sections and available decay channels. Previous works, for example 
\cite{Zheng:2016ygg}, have shown, for the case of non-interacting scalar DM, that old isolated neutron stars set strong bounds on the allowed scattering cross section between light quarks and DM. This could in practice strongly disfavour the accumulation of significant amounts of DM in NS through accretion. NSs with clouds could form in a similar way, given that either the DM is the dominant contribution to the FBS or that the DM properties only allow low-compactness configurations (e.g., when the particle mass is small \cite{Diedrichs:2023trk}). The fermionic and bosonic components could conceivably be separated from one another, e.g., during a supernova NS-kick \cite{Arzoumanian:2001dv,Holland-Ashford:2017tqz,Bezares:2022obu,Kulkarni:2020fhv}. There, the stellar remnant gets ejected and rapidly moves away from the remaining stellar envelope. This process could allow for NSs with a large range of possible DM-fractions. The DM particles most interesting for FBSs are generally (self-interacting) ultralight DM particles, weakly interacting massive particles, dark photons \cite{Biswas:2022tcw,An:2023mvf} (as a candidate for vector DM) and axions \cite{Diedrichs:2023trk,Huang:2018pbu,Zhang:2021mks,Baryakhtar:2022hbu,Bertone:2018krk,Peccei:1977ur,Peccei:1977hh,ADMX:2009iij,Galanti:2022ijh,Klasen:2015uma,Irastorza:2018dyq}. \\

Another formation channel is motivated through theories of modified gravity. One way of producing large amounts of scalar (or vector) fields is superradiance \cite{Brito:2015oca,Siemonsen:2022ivj}. Spontaneous scalarization \cite{Liebling:2012fv,Silva:2017uqg} also provides a way of producing significant scalar \cite{Liebling:2012fv,Damour:1996ke} and vector\footnote{In the case of vector fields, the process is also called spontaneous vectorization.} \cite{Silva:2021jya,Ramazanoglu:2017xbl} field amplitudes. It has also been studied explicitly in NSs \cite{Silva:2017uqg,Kase:2020yhw,Minamitsuji:2020pak} and could be a way of forming systems with scalar and vector fields. Scalarization might also take place dynamically in the late stages of the evolution of binary NS systems \cite{Barausse:2012da}, forming either a black hole or a FBS after merger (depending, e.g., on the initial masses of the binary objects). \\


Self gravitating vector fields have already been investigated. These objects are called Proca stars. They are modelled by a complex vector field and were first proposed by \cite{Brito:2015pxa}. They can be thought of as macroscopic condensates of spin-1 particles \cite{Liebling:2012fv}.  Proca stars have been studied by a number of groups analytically \cite{Cardoso:2021ehg,Aoki:2022woy,Brihaye:2017inn} and numerically \cite{Sanchis-Gual:2017bhw,Wang:2023tly}, such as in merger simulations \cite{CalderonBustillo:2020fyi,Sanchis-Gual:2018oui}. Different types of Proca stars with charge \cite{SalazarLandea:2016bys}, rotation \cite{Brito:2015pxa} and with a quartic self-interaction potential \cite{Minamitsuji:2018kof} were also considered. Other works \cite{Rosa:2022tfv,Herdeiro:2021lwl,Rosa:2022toh} studied shadow images of Proca stars in different scenarios. \\

In this work, we study the combined system of a vector field and NS matter, which we call fermion Proca stars (FPSs). Starting with an action for complex vector fields coupled minimally to gravity and nuclear matter, we derive a system of differential equations and solve them numerically (\autoref{subsec:Equilibrium_Solutions}). We also pedagogically motivate the boundary conditions (\autoref{subsec:Initial_Conditions}), find an analytical bound for the vector field amplitude and derive scaling relations in the equations of motion (\autoref{subsec:Analytical_Results}). The equations are solved using a shooting method and the integrator implemented in our code (for the code, see \cite{Diedrichs-Becker-Jockel}). The numerical methods are also explained in \autoref{subsec:Numerical_Methods}. We show radial profiles of FPSs (\autoref{subsec:Radial_Profiles}) and then compute global quantities such as mass and radius and compare them to astrophysical observations (\autoref{subsec:Stable_Solutions}). In \autoref{subsec:Comparison_Scalar_FBS}, we compare FPSs to their counterpart with a scalar field. In the following, we refer to the scalar case as ''fermion boson stars'' (FBSs). Finally, we compute higher modes of FPSs and compute configurations with different EOS (\autoref{subsec:Higher_Modes_and_different_EOS}). \\

We find that the vector field significantly affects the NS properties and thus produces detectable signatures. FPSs admit DM core and cloud solutions. 
Small DM masses lead to DM clouds, and large masses form DM cores. Core solutions compactify the NS component. Cloud solutions lead to less compact configurations. Some solutions appear to violate the Buchdahl limit when only observing the NS component. \\
We then compare FPSs (with a vector field) to FBSs (with a scalar field). FPSs tend to be more massive and geometrically larger than FBSs for equal boson masses and self-interaction strengths. For a given measurement, this would favor larger vector DM masses (compared to scalar DM), because larger DM masses produce smaller and less massive objects. \\
We find a significant amount of degenerate solutions between different choices of FBSs, FPSs, the DM properties and the EOS. For different boson masses and DM-fractions, FPSs and FBSs can both be degenerate with each other and also be degenerate with pure NSs with a different EOS. Using scaling relations for pure boson stars and Proca stars, we show that FBSs and FPSs are virtually indistinguishable if the boson masses differ by a factor of $1.671$ and the DM has no self-interactions. We confirm the existence of FPSs in higher modes which are stable under linear radial perturbations. \\

Throughout this work, we use units where $G = c = M_\odot = 1$ (also see Appendix \ref{sec:appendix:units}). The Einstein summation convention for tensors is implied. This paper is based on the Master thesis of Cédric Jockel \cite{Jockel:2023qnq}.

%% file: theory.tex
\subsection{Equilibrium Solutions}
\label{subsec:Equilibrium_Solutions}

Fermion Proca stars (FPSs) are combined systems of fermions and vector bosons, which interact only gravitationally. They can be seen as a macroscopic Bose-Einstein condensate which coexists with a NS at the same point in space. We model FPSs using a relativistic fluid for the NS component and a complex vector field for the bosonic component. FPSs are described by the Einstein-Proca system minimally coupled to a matter term $\mathcal{L}_m$,
\begin{align}
S = \int \sqrt{-g} \left( \frac{R}{2\kappa} - \frac{1}{2} \tensor{F}{_\mu_\nu} \tensor{\bar{F}}{^\mu^\nu} - V(\tensor{A}{_\rho} \tensor{\bar{A}}{^\rho}) - \mathcal{L}_{m} \right) dx^4 \: , \label{eq:fermion-boson-stars:fermion-proca-stars:fermion-proca-star-action}
\end{align}
where $R$ is the Ricci curvature scalar, $g$ is the determinant of the spacetime metric $\tensor{g}{_\mu_\nu}$ and $\kappa = 8\pi G/c^4$ is a constant. The bar denotes complex conjugation. $\tensor{F}{_\mu_\nu} = \tensor{\nabla}{_\mu} \tensor{A}{_\nu} - \tensor{\nabla}{_\nu} \tensor{A}{_\mu}$ is the antisymmetric field strength tensor and $V(\tensor{A}{_\rho} \tensor{\bar{A}}{^\rho})$ is the vector field potential. The latter depends solely on the magnitude of the vector field $\tensor{A}{_\rho} \tensor{\bar{A}}{^\rho}$. \\
By taking the variation of \eqref{eq:fermion-boson-stars:fermion-proca-stars:fermion-proca-star-action} with respect to the inverse spacetime metric $\delta \tensor{g}{^\mu^\nu}$, one obtains the Einstein equations
\begin{align}
\tensor{G}{_\mu_\nu} =& \, \kappa \left( T_{\mu \nu}^{(NS)} + T_{\mu \nu}^{(A)} \right) \: , \label{eq:fermion-boson-stars:fermion-proca-stars:einstein-equations-fps}
\end{align}
where $T_{\mu \nu}^{(NS)}$ and $T_{\mu \nu}^{(A)}$ are the energy-momentum tensors describing the NS matter and the vector field matter, respectively. The energy-momentum tensor of the NS matter is taken to be that of a perfect fluid:
\begin{align}
	T_{\mu \nu}^{(NS)} = (e + P) \tensor{u}{_\mu} \tensor{u}{_\nu} + P \tensor{g}{_\mu_\nu} \: . \label{eq:fermion-boson-stars:scalar-fermion-boson-stars:energy-momentum-tensor-perfect-fluid}
\end{align}
$P$ and $e$ are the pressure and the energy density of the fluid, respectively. The energy density $e$ is related to the rest mass density $\rho$ through $e=\rho(1+\epsilon)$, where $\epsilon$ is the internal energy. $\tensor{u}{_\mu}$ is the four-velocity of the fluid. The energy-momentum tensor \eqref{eq:fermion-boson-stars:scalar-fermion-boson-stars:energy-momentum-tensor-perfect-fluid} and the fluid flow $\tensor{J}{^\mu} := \rho \tensor{u}{^\mu}$ are conserved (implying conservation of energy-momentum and of the rest mass, respectively). This leads to the conservation equations
\begin{align}
	\tensor{\nabla}{_\mu} T^{\mu \nu}_{(NS)} = 0 \:\: , \:\: \tensor{\nabla}{_\mu} J^{\mu}_{\phantom{()}} = 0  \: . \label{eq:fermion-boson-stars:scalar-fermion-boson-stars:energy-momentum-and-restmass-conservation}
\end{align}
The conservation of the fluid flow $\tensor{J}{^\mu}$ allows us to define the conserved total rest mass of neutron matter, which we call the fermion number $N_\mathrm{f}$. We obtain the fermion number by integrating the right part of \eqref{eq:fermion-boson-stars:scalar-fermion-boson-stars:energy-momentum-and-restmass-conservation} over space,
\begin{align}
	N_\mathrm{f} := \int \sqrt{-g}\; \tensor{g}{^t^\mu} \tensor{J}{_\mu} dx^3 \: . \label{eq:fermion-boson-stars:scalar-fermion-boson-stars:definition-NS-restmass}
\end{align}
The energy-momentum tensor of the vector field is given by
\begin{align}
T_{\mu \nu}^{(A)} &= \tensor{F}{_\mu_\rho} \tensor{\bar{F}}{_\nu^\rho} + \tensor{\bar{F}}{_\mu_\rho} \tensor{F}{_\nu^\rho}  - \frac{1}{2} \tensor{g}{_\mu_\nu} \tensor{F}{^\rho^\sigma} \tensor{\bar{F}}{_\rho_\sigma} \label{eq:fermion-boson-stars:fermion-proca-stars:energy-momentum-tensor-vector-field} \\
&+ \tensor{g}{_\mu_\nu} V(\tensor{A}{_\rho} \tensor{\bar{A}}{^\rho}) + V'(\tensor{A}{_\rho} \tensor{\bar{A}}{^\rho}) ( \tensor{A}{_\mu} \tensor{\bar{A}}{_\nu} + \tensor{A}{_\nu} \tensor{\bar{A}}{_\mu} ) \: , \nonumber 
\end{align}
where the derivative of the potential $V$ is
\begin{align}
 V'(A_\rho \bar{A}^\rho ) := \frac{d V(A_\rho \bar{A}^\rho ) }{d (A_\rho \bar{A}^\rho)}  \: . \label{eq:fermion-boson-stars:fermion-proca-stars:derivative-of-vector-potential}
\end{align}
The equations of motion (Proca equations) of the vector field and the complex conjugate are computed from the action \eqref{eq:fermion-boson-stars:fermion-proca-stars:fermion-proca-star-action} using the Euler-Lagrange equations for a complex vector field. One obtains
\begin{align}
\tensor{\nabla}{^\mu} \tensor{\bar{F}}{_\mu_\nu} = V'(\tensor{A}{_\rho} \tensor{\bar{A}}{^\rho} ) \tensor{\bar{A}}{_\nu} \: , \:  \tensor{\nabla}{^\mu} \tensor{F}{_\mu_\nu} = V'(\tensor{A}{_\rho} \tensor{\bar{A}}{^\rho} ) \tensor{A}{_\nu} \: . \label{eq:fermion-boson-stars:fermion-proca-stars:proca-equations}
\end{align} 
The covariant derivative of \eqref{eq:fermion-boson-stars:fermion-proca-stars:proca-equations} is zero, i.e., $\tensor{\nabla}{^\mu} \tensor{\nabla}{^\nu} \tensor{F}{_\mu_\nu} = 0$. This leads to a dynamical constraint on the field derivative, resembling the Lorentz condition used in the Maxwell and Proca equations (also see \cite{Liebling:2012fv,Brito:2015pxa}):
\begin{align}
\tensor{\nabla}{^\nu} \tensor{A}{_\nu} = - \frac{\tensor{\nabla}{^\nu} \left[ V'(\tensor{A}{_\rho} \tensor{\bar{A}}{^\rho} ) \right]}{ V'(\tensor{A}{_\rho} \tensor{\bar{A}}{^\rho} ) } \tensor{A}{_\nu} \: . \label{eq:fermion-boson-stars:fermion-proca-stars:proca-constraint}
\end{align}
This constraint could be useful in numerical simulations to track the numerical error and assess constraint violations of a given numerical scheme. The global $U(1)$-symmetry in the Lagrangian \eqref{eq:fermion-boson-stars:fermion-proca-stars:fermion-proca-star-action} under the transformation of the vector field $\tensor{A}{_\mu}$ (and $ \tensor{\bar{A}}{_\mu}$) gives rise to a conserved Noether current
\begin{align}
j^\mu = i \left( \bar{F}^{\mu\nu} A_\nu - F^{\mu\nu} \bar{A}_\nu \right) \: . \label{eq:fermion-boson-stars:fermion-proca-stars:noether-current}
\end{align}
The conserved quantity (i.e., the Noether charge) associated to \eqref{eq:fermion-boson-stars:fermion-proca-stars:noether-current} is obtained by integrating the conservation equation $\nabla_\mu j^\mu = 0$ over space,
\begin{align}
N_\mathrm{b} := \int \sqrt{-g} g^{t\mu} j_\mu dx^3 \: . \label{eq:fermion-boson-stars:scalar-fermion-boson-stars:conserved-boson-number-definition}
\end{align}
$N_\mathrm{b}$ is called the boson number and is related to the total number of bosons present in the system. It can equivalently also be interpreted as the total rest mass energy of the bosonic component of the FPS. \\

We proceed by solving the Einstein equations \eqref{eq:fermion-boson-stars:fermion-proca-stars:einstein-equations-fps} and the Proca equations \eqref{eq:fermion-boson-stars:fermion-proca-stars:proca-equations} for spherically symmetric and static configurations in equilibrium. For that, we consider the spherically symmetric ansatz for the spacetime metric
\begin{align}
\tensor{g}{_\mu_\nu} = \text{diag} \left( - \alpha^2(r), \: a^2(r), \: r^2, \: r^2\sin^2(\theta) \right) \: . \label{eq:fermion-boson-stars:fermion-proca-stars:metric-ansatz}
\end{align}
We further assume the perfect fluid to be static, such that the four-velocity can be written as
\begin{align}
\tensor{u}{^\mu} = \left(- \frac{1}{\alpha}, 0, 0, 0 \right) \: , \: \tensor{u}{_\mu} = (\alpha, 0, 0, 0) \: . \label{eq:fermion-boson-stars:fermion-proca-stars:four-velocity-static-fluid}
\end{align}
For the vector field, we employ the harmonic phase ansatz and a purely radial vector field (see \cite{Brito:2015pxa,Minamitsuji:2018kof,Herdeiro:2020kba, Brihaye:2017inn,SalazarLandea:2016bys}). The vector field is then given by
\begin{align}
A_\mu (t,x) = e^{-i \omega t} (E(r), iB(r), 0, 0)  \: , \label{eq:fermion-boson-stars:fermion-proca-stars:vector-field-harmonic-ansatz}
\end{align}
where $\omega$ is the vector field frequency and $E(r)$, $B(r)$ are purely radial real functions. \\
Using the spherical symmetric metric ansatz \eqref{eq:fermion-boson-stars:fermion-proca-stars:metric-ansatz} together with the harmonic phase ansatz \eqref{eq:fermion-boson-stars:fermion-proca-stars:vector-field-harmonic-ansatz} for the vector field, we solve the Einstein equations and obtain the equations of motion. 
One obtains an expression for the radial derivative of $a(r)$ by re-arranging the $tt$-component of \eqref{eq:fermion-boson-stars:fermion-proca-stars:einstein-equations-fps}. We then divide the $tt$- and $rr$-components of \eqref{eq:fermion-boson-stars:fermion-proca-stars:einstein-equations-fps} by $\alpha^2$ and $a^2$, respectively. We add both terms and find a direct relation between the first radial derivatives of $a(r)$ and $\alpha(r)$. We use this to solve for the derivative of $\alpha(r)$. \\
The evolution equations for the vector field components can be computed from the Proca equations \eqref{eq:fermion-boson-stars:fermion-proca-stars:proca-equations}. It does not matter which equation of \eqref{eq:fermion-boson-stars:fermion-proca-stars:proca-equations} is used since the complex phase will cancel out and will leave only the radial functions in both cases. The $\nu=r$ component yields the equation of motion for $E(r)$. The $\nu=t$ component of \eqref{eq:fermion-boson-stars:fermion-proca-stars:proca-equations} gives us the equation of motion for $B(r)$. Finally, the $r$-component of the conservation equation for the energy-momentum tensor (left side of \eqref{eq:fermion-boson-stars:scalar-fermion-boson-stars:energy-momentum-and-restmass-conservation}) provides a differential equation for the pressure $P(r)$. For a more detailed derivation, we refer to \cite{Jockel:2023qnq}. The full equations of motion for the Einstein-Proca system coupled to matter are thus:
\begin{widetext}
\begin{subequations} 
\begin{align}
a' = \frac{da}{dr} &=  \frac{a}{2} \left[ \frac{(1-a^2)}{r} + 8\pi r a^2 \left( \:\: e + \frac{1}{\alpha^2 a^2} (E' - \omega B)^2 + V(A_\rho \bar{A}^\rho ) + 2 V'(A_\rho \bar{A}^\rho ) \frac{E^2}{\alpha^2}  \right)  \right] \: , \label{eq:fermion-boson-stars:fermion-proca-stars:TOV-equations-grr} \\
\alpha' = \frac{d \alpha}{dr} &= \frac{\alpha}{2} \left[ \frac{(a^2 -1)}{r} + 8\pi r a^2 \left( P - \frac{1}{\alpha^2 a^2} (E' - \omega B)^2 - V(A_\rho \bar{A}^\rho ) + 2 V'(A_\rho \bar{A}^\rho ) \frac{B^2}{a^2}  \right)  \right] \: , \label{eq:fermion-boson-stars:fermion-proca-stars:TOV-equations-gtt} \\
E'  = \frac{dE}{dr} &= - V'(A_\rho \bar{A}^\rho ) \frac{B \alpha^2}{\omega} + \omega B  \: , \label{eq:fermion-boson-stars:fermion-proca-stars:TOV-equations-E} \\
B' = \frac{dB}{dr} &= \left\{ V''(A_\rho \bar{A}^\rho ) \left( \frac{2B^2 a'}{a^3} + \frac{2E E'}{\alpha^2} - \frac{2E^2 \alpha'}{\alpha^3} \right) \frac{B \alpha^2}{\omega} - V'(A_\rho \bar{A}^\rho ) \left( a^2 E + \frac{2 B \alpha \alpha'}{\omega} \right) \right. \nonumber \\
&- \left. \left( \frac{a'}{a\,} + \frac{\alpha'}{\alpha\,} - \frac{2}{r} \right) (E' - \omega B) \right\} \left( V''(A_\rho \bar{A}^\rho ) \frac{2}{\omega} \frac{B^2 \alpha^2}{a^2} + V'(A_\rho \bar{A}^\rho ) \frac{\alpha^2}{\omega} \right)^{-1} \: , \label{eq:fermion-boson-stars:fermion-proca-stars:TOV-equations-B} \\
P' =\frac{d P}{dr} &= -\left[ e + P \right] \frac{\alpha'}{\alpha} \: . \label{eq:fermion-boson-stars:fermion-proca-stars:TOV-equations-P}
\end{align}
\end{subequations}
\end{widetext}

This system of equations is closed by providing an equation of state $P(e)$ (or $P(\rho,\epsilon)$) for the nuclear matter part. \\
Note that all equations are first-order differential equations. This is different to scalar FBSs where an additional variable has to be introduced to make the system first-order (see, e.g., \cite{Diedrichs:2023trk,DiGiovanni:2021ejn}). Another difference is that no derivative of the potential enters the equations of motion for the metric components in the scalar field case, but it does for the vector field case. \\
For the considered system and ansatz for the metric \eqref{eq:fermion-boson-stars:fermion-proca-stars:metric-ansatz} and vector field \eqref{eq:fermion-boson-stars:fermion-proca-stars:vector-field-harmonic-ansatz}, the expressions for the fermion number \eqref{eq:fermion-boson-stars:scalar-fermion-boson-stars:definition-NS-restmass} and boson number \eqref{eq:fermion-boson-stars:scalar-fermion-boson-stars:conserved-boson-number-definition} simplify to
\begin{subequations}
\begin{align}
N_\mathrm{f} &= 4\pi \int_0^{R_\mathrm{f}} a \rho r^2 dr \: , \label{eq:fermion-boson-stars:fermion-proca-stars:conserved-fermion-number-simplified} \\
N_\mathrm{b} &= 8 \pi \int_0^\infty B \frac{(\omega B - E')}{\alpha a} r^2 dr \: . \label{eq:fermion-boson-stars:fermion-proca-stars:conserved-boson-number-simplified}
\end{align}
\end{subequations}
$R_\mathrm{f}$ denotes the fermionic radius (i.e., the radius of the NS component). It is defined by the radial position at which the pressure $P$ of the NS component reaches zero. It is also possible to define the bosonic radius $R_\mathrm{b}$. It is defined as the radius where $99\,\%$ of the bosonic restmass $N_\mathrm{b}$ (see \eqref{eq:fermion-boson-stars:fermion-proca-stars:conserved-boson-number-simplified}) is contained. Using these definitions we gain the ability to discriminate between DM core and cloud solutions. Core solutions have $R_\mathrm{f} > R_\mathrm{b}$ and cloud solutions have $R_\mathrm{f} < R_\mathrm{b}$. The total gravitational mass is defined in the limit of large radii, imposing that the solution asymptotically converges to the Schwarzschild solution
\begin{align}
	M_\mathrm{tot} := \lim_{r \rightarrow \infty} \frac{r}{2} \left( 1 - \frac{1}{(a(r))^2 } \right) \: . \label{eq:fermion-boson-stars:fermion-proca-stars:definition-BS-total-gravitational-mass}
\end{align}

\subsection{Initial Conditions}
\label{subsec:Initial_Conditions}

We derive the boundary conditions of equations \eqref{eq:fermion-boson-stars:fermion-proca-stars:TOV-equations-grr}-\eqref{eq:fermion-boson-stars:fermion-proca-stars:TOV-equations-P} at $r = 0$ and at $r= \infty$. The values at the origin will later serve as initial conditions for the numerical integration. We first consider the equations of motion in the limit $r \rightarrow 0$ while imposing regularity at the origin (i.e., the solution must not diverge). We first analyze equation \eqref{eq:fermion-boson-stars:fermion-proca-stars:TOV-equations-grr}. The term proportional to $1/r$ dominates at small radii and will diverge if $r \rightarrow 0$. Thus, the only way to maintain regularity is to set $a(r=0) = 1$. It directly follows that $a'(r=0) = 0$. Similarly, equation \eqref{eq:fermion-boson-stars:fermion-proca-stars:TOV-equations-gtt} leads to $\alpha'(r=0) = 0$. The exact value of $\alpha (r=0) = \alpha_0$ is a priori undetermined and can be chosen in a way thought suitable. We will elaborate on this in \autoref{subsec:Initial_Conditions}. \\
The initial conditions for the vector field components $E(r)$ and $B(r)$ can be obtained in a similar manner. We first consider \eqref{eq:fermion-boson-stars:fermion-proca-stars:TOV-equations-B}. In the limit $r\rightarrow 0$, the term proportional to $1/r$ dominates and regularity then demands that $E' = \omega B$. It follows that $B'(r=0) = 0$. This result can be inserted into \eqref{eq:fermion-boson-stars:fermion-proca-stars:TOV-equations-E}, which leads to the relation
\begin{align}
    \begin{split}
     E'  =& \,\omega B = - V'(A_\rho \bar{A}^\rho ) \frac{B \alpha^2}{\omega} + \omega B \\ 
     &\Longrightarrow \: 0 = V'(A_\rho \bar{A}^\rho ) B \alpha^2\: .
    \end{split}
\end{align}
Since at $r=0$, $\alpha(r=0) \neq 0$ and $V' \neq 0$ in general, this relation can only be fulfilled if we demand that $B(r=0) = 0$. Plugging this relation into \eqref{eq:fermion-boson-stars:fermion-proca-stars:TOV-equations-E} yields $E'(r=0) = 0$. The central value of the field $E'(r=0) = E_0$ is therefore undetermined by the equations of motion, and thus is a free parameter of the theory. \\
A similar analysis at large distances reveals the boundary conditions at $r \rightarrow \infty$ for all variables. We impose an asymptotically flat spacetime. This requires that $a(r \rightarrow \infty) = \alpha(r \rightarrow \infty) = 1$. All terms proportional to $r$ in \eqref{eq:fermion-boson-stars:fermion-proca-stars:TOV-equations-grr} and \eqref{eq:fermion-boson-stars:fermion-proca-stars:TOV-equations-gtt} must vanish at infinity to fulfill the flat-spacetime limit. Therefore, the vector field components must vanish at infinity, $E(r \rightarrow \infty) = 0$ and $B(r \rightarrow \infty) = 0$. Pressure $P(r)$, energy density $e(r)$ and rest mass density $\rho$ must be zero outside the NS component of the FPS. This will happen at the fermionic radius $R_\mathrm{f}$. We summarize all boundary conditions in the following:
\begin{align}
    \begin{split}
        \lim_{r \rightarrow \infty} a(r) &= 1 \:\: , \:\: a(0) = 1 \: , \\
        \lim_{r \rightarrow \infty} \alpha(r) &= 1 \:\: , \:\: \alpha(0) = \alpha_0 \: , \\
        \lim_{r \rightarrow \infty} E(r) &= 0 \:\: , \:\: E(0) = E_0 \: , \\
        \lim_{r \rightarrow \infty} B(r) &= 0 \:\: , \:\: B(0) = 0 \: , \\
        \rho(r > R_\mathrm{f}) &= 0 \:\: , \:\: \rho(0) = \rho_c \: .
    \end{split} \label{eq:fermion-boson-stars:fermion-proca-stars:TOV-initial-conditions}
\end{align}

The initial condition for the metric component $\alpha(0) = \alpha_0$ is fixed by its behavior at infinity. \\

We also note a popular and widely employed alternative to the self-consistent wave-treatment of DM performed in this work -- the two-fluid formalism (see e.g. \cite{Karkevandi:2021ygv,Leung:2022wcf,Xiang:2013xwa}). In there, the nuclear matter and the DM are both modeled as perfect fluids, which only interact gravitationally. One can then solve the Einstein equations and obtain a set of modified TOV equations that describe the density distribution of both fluids. Although simplistic, this formalism has the advantage that it is easy to implement, numerically cheap and is applicable to a wide range of fermionic and bosonic DM models. It is also possible to use arbitrary effective EOS for the dark matter. A disadvantage of this model is that it ignores possible wave properties of ultralight DM, that we study in this work. It is also only possible to study non-excited ground states of the wave-like DM. Further, some emerging properties like the maximal bound on the vector field amplitude (see \eqref{eq:fermion-boson-stars:fermion-proca-stars:analytical-bound-amplitude}) would not be captured by the two-fluid formalism.

\subsection{Analytical Results}
\label{subsec:Analytical_Results}

For a scalar (fermion) boson star, one can scale the field frequency $\omega$ to absorb the initial value of $\alpha_0$ so that it may be set to one (see, e.g., \cite{Diedrichs:2023trk}). We investigate whether a similar scaling relation also exists for FPSs. We find that the equations of motion \eqref{eq:fermion-boson-stars:fermion-proca-stars:TOV-equations-grr}-\eqref{eq:fermion-boson-stars:fermion-proca-stars:TOV-equations-P} are invariant when simultaneously scaling the following variables as
\begin{align}
	\tilde{\alpha} = \sigma \alpha \:\: , \:\: \tilde{\omega} = \sigma \omega \:\: , \:\: \tilde{E} = \sigma E \:\: , \:\: \text{where} \:\: \sigma \in \mathbb{R} \: . \label{eq:fermion-boson-stars:fermion-proca-stars:TOV-scaling-relations}
\end{align}
The potential $V(A_\rho \bar{A}^\rho)$ is always invariant with respect to this scaling because
\begin{align}
	A_\rho \bar{A}^\rho = \left( \frac{B^2}{a^2} - \frac{E^2}{\alpha^2} \right) = \left( \frac{B^2}{a^2} - \frac{\tilde{E}^2}{\tilde{\alpha}^2} \right) \: .
\end{align}
The invariance of \eqref{eq:fermion-boson-stars:fermion-proca-stars:TOV-equations-grr}-\eqref{eq:fermion-boson-stars:fermion-proca-stars:TOV-equations-P} under the scaling relation \eqref{eq:fermion-boson-stars:fermion-proca-stars:TOV-scaling-relations} thus allows us to choose $\sigma$ in such a way that the initial condition for $\alpha(0) = \alpha_0$ may be set to $\alpha_0 = 1$ \footnote{Or one could, in principle, also re-scale $E_0$ to always be equal to one.}. We will make use of this relation in the numerical analysis. All pre-scaling physical values can be recovered from the asymptotic behavior of $\alpha(r \rightarrow \infty)$ by performing the inverse transformation to \eqref{eq:fermion-boson-stars:fermion-proca-stars:TOV-scaling-relations}. Note that the expression for total gravitational mass \eqref{eq:fermion-boson-stars:fermion-proca-stars:definition-BS-total-gravitational-mass} is not affected by this scaling. \\
In contrast to the scaling relation of boson stars with a scalar field, where only the frequency $\omega$ and the metric component $\alpha$ are re-scaled, the vector field component $E$ is also affected in the case of Proca stars. To our knowledge, this is the first time the scaling relation \eqref{eq:fermion-boson-stars:fermion-proca-stars:TOV-scaling-relations} has been mentioned explicitly (apart from the Master thesis \cite{Jockel:2023qnq} which precedes this work). \cite{Minamitsuji:2018kof} briefly mentioned scaling the frequency but not the vector field component. \\

We also report an analytical bound on the central vector field amplitude $E(0)=E_0$. Equations \eqref{eq:fermion-boson-stars:fermion-proca-stars:TOV-equations-E} and \eqref{eq:fermion-boson-stars:fermion-proca-stars:TOV-equations-B} govern the dynamics of the vector field. Note that the term in the denominator of the equation of motion for $B(r)$ \eqref{eq:fermion-boson-stars:fermion-proca-stars:TOV-equations-B} could in some cases lead to singularities. We analyze the behavior of the denominator by setting it equal to zero. This leads to a remarkable behavior when considering a quartic self-interaction potential $V$ of the form
\begin{align}
  V(A_\mu \bar{A}^\mu ) = m^2 A_\mu \bar{A}^\mu + \frac{\lambda}{2} ( A_\mu \bar{A}^\mu )^2 \: , \label{eq:fermion-boson-stars:fermion-proca-stars:quartic-self-interaction-potential}
\end{align}
where $m$ is the mass of the vector boson and $\lambda$ is the self-interaction parameter. We insert the potential \eqref{eq:fermion-boson-stars:fermion-proca-stars:quartic-self-interaction-potential} into the singular term in \eqref{eq:fermion-boson-stars:fermion-proca-stars:TOV-equations-B} and obtain
\begin{align}
 \left( \frac{E^2}{\alpha^2} - \frac{3B^2}{a^2} \right) =\frac{m^2}{\lambda} \: . \label{eq:fermion-boson-stars:fermion-proca-stars:B-equation-singularity-criterium-rewritten}
\end{align}
This expression holds for all radii. We analyze its behavior in the limit $r \rightarrow 0$ by applying the initial conditions given in \eqref{eq:fermion-boson-stars:fermion-proca-stars:TOV-initial-conditions}. One obtains a critical value for the central field amplitude $E_0$:
\begin{align}
E_{0,\text{crit}} = \frac{m \alpha_0}{\sqrt{\lambda}} = \frac{\alpha_0}{\sqrt{ 8\pi \Lambda_{\mathrm{int}}} } \: . \label{eq:fermion-boson-stars:fermion-proca-stars:analytical-bound-amplitude}
\end{align}
We here also defined the dimensionless interaction parameter $\Lambda_{\mathrm{int}} = \lambda / 8 \pi m^2$. This expression constitutes an analytical upper bound for the central amplitude of the vector field. This means that any FPS with initial conditions for the field larger than $E_{0,\text{crit}}$ will be physically forbidden, since \eqref{eq:fermion-boson-stars:fermion-proca-stars:TOV-equations-B} will become singular and diverge. This result matches the analytical bound found by \cite{Minamitsuji:2018kof}. \\
The relation implies that for strong self-interaction strengths $\Lambda_{\mathrm{int}}$, the allowed range for Proca stars becomes increasingly small and vanishes in the limit of very strong self-interactions. This fact could conceivably be used to constrain the vector field parameters $m$ and $\lambda$. For example, a maximal vector field amplitude implies a maximal amount of accretion of vector bosons until the system becomes unstable. The field would then either dissipate to infinity, shed the excess vector field component, or collapse into a black hole. We leave a thorough investigation for future work.

\subsection{Stability Criterion}
\label{subsec:Stability_Criterion}

Every FPS solution is characterized by the initial values for the central density $\rho_c$ and the central value of the vector field $E_0$. When studying them in astrophysical contexts, the question of stability of FPSs naturally arises. The stability of pure Proca stars and NSs to radial perturbations is well known (see \cite{Brito:2015pxa} for Proca stars). The stable and unstable solutions are separated by the point at which the total gravitational mass reaches its maximum with regard to the central density $\rho_c$ (for NS) and the central field $E_0$ (for Proca stars). \\
Since FPSs are two-parameter solutions, the stability criterion needs to be modified. It was first presented for scalar FBSs by \cite{Henriques:1990xg} (also see \cite{Liebling:2012fv} for a review). But the criterion is more general and can also be applied to systems of two gravitationally interacting fluids. This is why we apply it here for FPSs. \\
The idea behind the generalized stability criterion is to find extrema in the total number of particles (fermion number $N_\mathrm{f}$ or boson number $N_\mathrm{b}$) for a fixed total gravitational mass. The transition between stable and unstable configurations is given by the point at which
\begin{align}
    \frac{d N_\mathrm{f}}{d \sigma} = \frac{d N_\mathrm{b}}{d \sigma} = 0 \: , \label{eq:fermion-boson-stars:scalar-fermion-boson-stars:FBS-stability-criterion}
\end{align}
where $d/d\sigma$ denotes the derivative in the direction of constant total gravitational mass (see \cite{Henriques:1990xg}). Up to a normalization factor, \eqref{eq:fermion-boson-stars:scalar-fermion-boson-stars:FBS-stability-criterion} can be written as
\begin{equation}
    \frac{d N_\mathrm{f}}{d\sigma} \propto - \frac{\partial M_\text{tot}}{\partial \rho_c} \frac{\partial N_\mathrm{f}}{\partial E_0} + \frac{\partial M_\text{tot}}{\partial E_0} \frac{\partial N_\mathrm{f}}{\partial \rho_c} \: . \label{eq:fermion-boson-stars:scalar-fermion-boson-stars:FBS-stability-criterion-rewritten}
\end{equation}
If one is only interested in the precise points where FPSs become unstable, the unspecified normalization factor in \eqref{eq:fermion-boson-stars:scalar-fermion-boson-stars:FBS-stability-criterion-rewritten} becomes irrelevant, since the whole relation is set to zero. \\
In summary, the stability criterion \eqref{eq:fermion-boson-stars:scalar-fermion-boson-stars:FBS-stability-criterion} can be used to discriminate between astrophysically stable and unstable FPS solutions. When perturbed, unstable solutions will either collapse to a black hole, dissipate to infinity or migrate to a stable solution through internal re-configuration (see \cite{Liebling:2012fv}).

\subsection{Numerical Methods}
\label{subsec:Numerical_Methods}

In this work, we solve the equations \eqref{eq:fermion-boson-stars:fermion-proca-stars:TOV-equations-grr}-\eqref{eq:fermion-boson-stars:fermion-proca-stars:TOV-equations-P} numerically to obtain self-consistent FPS solutions. We have implemented the algorithm in the code \cite{Diedrichs-Becker-Jockel} which was developed by the authors of \cite{Diedrichs:2023trk}. The equations have one parameter undetermined by the boundary conditions \eqref{eq:fermion-boson-stars:fermion-proca-stars:TOV-initial-conditions}, namely the vector field frequency $\omega$. We use a shooting-algorithm to find $\omega$ numerically. For given $\rho_c$ and $E_0$, there exist only discrete values of $\omega$, such that the boundary conditions at infinity \eqref{eq:fermion-boson-stars:fermion-proca-stars:TOV-initial-conditions} are fulfilled. These discrete values are called eigenvalues or modes. There are infinitely many of these modes. They are characterized by the number of roots (i.e., zero-crossings) the field $E(r)$ has. Usually we are only interested in the lowest mode, since only it is believed to be dynamically stable \cite{Liebling:2012fv}. The lowest mode of the vector field always has one root in $E(r)$. The following algorithm can however be used to find any desired mode. \\

We integrate the system of ordinary differential equations \eqref{eq:fermion-boson-stars:fermion-proca-stars:TOV-equations-grr}-\eqref{eq:fermion-boson-stars:fermion-proca-stars:TOV-equations-P} using a fifth order accurate Runge-Kutta-Fehlberg solver for some fixed value of $\omega$. The vector field will then diverge towards positive or negative infinity at some finite radius. The system only converges at infinity if any mode is hit directly. But this is impossible to achieve numerically with finite precision. We thus make use of this diverging property to find the wanted frequency mode. When the frequency $\omega$ is close to the wanted mode, the divergence will happen at increasingly large radii, the closer the chosen value for $\omega$ is to the mode. A higher accuracy in finding $\omega$ will therefore push the divergence to larger radii. When $\omega$ is not exactly tuned to the mode, the vector field profile $E(r)$ will diverge towards $+\infty$ or $-\infty$ and change its direction of divergence when $\omega$ passes a mode. The direction of divergence depends on which mode is solved for. For modes with an even number of roots, the field will diverge to $+\infty$ if the frequency $\omega$ is below the mode, and it will diverge to $-\infty$ if $\omega$ is above the mode. This will be reversed for all modes with an odd number of roots. By making use of the direction of divergence, we gain a binary criterion to find the correct mode. The value of $\omega$ can then be adapted -- increased or decreased -- based on the direction of divergence and the wanted mode. This procedure requires to integrate the system of equations multiple times with different values for $\omega$, until the correct value is found. \\
We implement this method in our code \cite{Diedrichs-Becker-Jockel} using a bisection algorithm, which converges exponentially fast. We start with upper and lower values of $\omega$, which are guaranteed to be smaller/larger than the wanted value of $\omega$ at the mode. In practice, lower and upper bounds of $\omega_\mathrm{bound} = [1,10]$ have proven to be numerically robust. We then perform the bisection search by taking the middle value of $\omega$ in this range and counting the number of roots in $E(r)$ at each step. This also allows us to discriminate between different modes and to target specific modes by demanding a certain number of roots in the field $E(r)$. The bisection is complete when the current value of $\omega$ found through bisection is close enough to the value of the mode. In our experience, the absolute accuracy needed to obtain robust solutions is on the order of $\Delta\omega = | \omega_\mathrm{mode} - \omega_\mathrm{bisection} | \approx 10^{-15}$. \\

Once a sufficiently accurate frequency $\omega$ is found, we modify the integration, such that $E(r)$ and $B(r)$ are set to zero at a finite radius $r_{B}^*$. This radius $r_{B}^*$ is defined at the point where the field $E(r)$ and its derivative $E'(r)$ are small. This roughly corresponds to the last minimum of $E(r)$ before it diverges. Also note that this is different to the bosonic radius $R_\mathrm{b}$ defined previously. The condition can be summarized as the point where $E(r_{B}^*)/ E_0 < 10^{-4}$ and $E'(r_{B}^*) \ll 1$. This is necessary because the interplay of the vector field and the NS matter can complicate the numerical solution. In some parts of the parameter space, especially for small initial densities $\rho_c$, the vector field could diverge while still inside the NS component, i.e., before the pressure $P(r)$ reaches zero (within numerical precision, we consider the pressure to be zero when $P < 10^{-15}$). This divergence would make finding physical values such as the fermionic radius $R_\mathrm{f}$ impossible. Therefore, we artificially set $E = B = 0$ for $r > r_{B}^*$. This allows us to circumvent the divergence and accurately resolve the rest of the NS component. Note that the divergence of the vector field only happens because it is impossible to perfectly tune $\omega$ to the exact value, within numerical precision. If $\omega$ could be found exactly, the divergence would not happen. Setting the vector field to zero at some radius $r_B^*$ is thus simply a way to maintain numerical stability of our algorithm. The condition was chosen so that the remaining contribution of the vector field to the other quantities (i.e., the metric components) is minimized. We have tested this method for different thresholds and confirmed that all extracted results are the same. \\

After integrating the solution to radii outside the matter sources (i.e. where $E=B=P=0$), we can extract global observables such as the total gravitational mass and radius. The outside of the source is located at radii $r$ larger than both the fermionic radius $R_\mathrm{f}$ and $r_{B}^*$. In this regime, neither the NS matter nor the vector field contribute significantly. There, we can extract the total gravitational mass $M_\mathrm{tot}$ \eqref{eq:fermion-boson-stars:fermion-proca-stars:definition-BS-total-gravitational-mass} and then compute the integrals \eqref{eq:fermion-boson-stars:fermion-proca-stars:conserved-fermion-number-simplified} and \eqref{eq:fermion-boson-stars:fermion-proca-stars:conserved-boson-number-simplified} to obtain the fermion/boson numbers $N_\mathrm{f}$, $N_\mathrm{b}$. \\
The vector field convergence condition $E(r_{B}^*)/ E_0 < 10^{-4}$ cannot be fulfilled for some configurations due to numerical precision limits. This generally happens for small initial field values $E_0 \lesssim 10^{-4}$, where the vector field extends far outside the NS component. In these cases, we extract the total gravitational mass $M_\text{tot} = \frac{1}{2} r_\mathrm{ext} (1 - a^{-2}(r_\mathrm{ext}))$ at the point where its derivative has a global minimum. When the vector field diverges, also the metric components do, and with it also $M_\text{tot}$. By taking the point where the derivative of the mass has a global minimum, which roughly corresponds to where the vector field and its derivative is closest to zero, we get the best possible estimate of the mass of the system before the divergence. \\

During our numerical analysis, we encountered the phenomenon that the bisection algorithm to find the frequency $\omega$ could fail for some specific initial conditions for $E_0$ and $\rho_c$. We found this to be the case due to the bisection algorithm jumping over multiple modes in one iteration step. The wanted mode was then skipped and ended up outside the bisection bounds. The bisection then converged on an unwanted $\omega$-value, or ended up failing entirely. We solved this problem by employing a backup algorithm that activates if the bisection fails. It restarts the bisection for $\omega$ but with different lower and upper bounds of $\omega_\mathrm{bound}$. We tested the backup algorithm for $4800$ FPS configurations with different vector field masses $m$ and self-interaction strengths $\Lambda_{\mathrm{int}} = \lambda/8\pi m^2$ with equally distributed initial conditions for $E_0$ and $\rho_c$. We found that $330$ ($\approx 6.8\,\%$) of all configurations needed one restart of the bisection, and only $3$ ($\approx 0.06\,\%$) of all configurations needed two restarts. In none of the tested cases, the bisection had to be restarted three times or more.

%% file: results.tex
We consider FPSs with a quartic self-interaction potential of the same form as in \eqref{eq:fermion-boson-stars:fermion-proca-stars:quartic-self-interaction-potential}. We further define the effective self-interaction parameter $\Lambda_\mathrm{int} = \lambda / 8\pi m^2$. The parameter $\Lambda_\mathrm{int}$ is a useful measure for the self-interaction strength and parametrizes scaling relations for the total gravitational mass $M_\mathrm{max} \approx  1.058 M^2_p / m$ \cite{Brito:2015pxa} (for small $\Lambda_\mathrm{int}$) and $M_\mathrm{max} \approx \sqrt{\Lambda_\mathrm{int}} \ln(\Lambda_\mathrm{int})\, M^2_p / m$ \cite{Minamitsuji:2018kof} (for large $\Lambda_\mathrm{int}$). These scaling relations and numerical pre-factors can be derived numerically by fitting the configurations of maximum mass for a given $m$ and $\Lambda_\mathrm{int}$ to postulated scaling behaviors. Note that the parameter $\Lambda_\mathrm{int}$ was originally introduced in the context of pure Proca stars and thus the scaling relations will not be generally valid for the mixed system. They can however be useful to understand the limiting cases where the FPS is dominated by the bosonic component. Nonetheless, we regard $\Lambda_\mathrm{int}$ to be a useful measure to compare different choices of the mass and self-interaction strength. The self-interaction parameter $\Lambda_\mathrm{int}$ in our work differs from the one used in \cite{Minamitsuji:2018kof} by a factor of two, even though they are defined in the same way. This is because a different normalization was used for the vector field. \\

We hereafter investigate models with parameters in the order of $m \approx 1.34 \e{-10}\,eV$ and $\Lambda_\mathrm{int} \approx 0-100$. This mass range is chosen because in this work we want to study DM which behaves as a macroscopic wave on typical neutron star length scales. Therefore, the correlation length of the ultralight DM particle is on the scale of $km$. Due to our units of $c=G=M_\odot=1$, lengths are measured in units of half the Schwarzschild radius of the Sun ($\approx1.48\,km$). Then the reduced Compton wavelength of the bosonic field is also measured in these units. $m=1$ in our code units thus corresponds to $1.336 \e{-10}\,eV$ (see a detailed explanation in Appendix \ref{sec:appendix:units}). The range for the self-interaction parameter was chosen so that it fulfills the observational constraints for the DM cross-section of $1\,cm^2/g$ obtained from the Bullet Cluster \cite{Eby:2015hsq,Sagunski:2020spe}. The choice of $\Lambda_\mathrm{int}$ is thus consistent with observations as long as it fulfills:
\begin{align}
    \pi \Lambda_\mathrm{int}^2 m  &= \frac{\lambda^2}{64 \pi m^3 } = \frac{1}{4} \frac{\sigma}{m} \stackrel{!}{<} 1 \frac{\text{cm}^2}{\text{g}} \:\:\:\:\:\: \iff  \\ 
    \Lambda_\mathrm{int} &\stackrel{!}{<} 8.5 \e{25} \left( \frac{\sigma/m}{ cm^2/g} \right)^{\frac{1}{2}} \left(\frac{1.34 \e{-10}\,eV}{m} \right)^{\frac{1}{2}} \: . \nonumber
\end{align}
Note that, in our conventions, the units of $\Lambda_\mathrm{int}$ are [solar gravitational radius] divided by [$1.336 \e{-10}\,eV$]. We used that the total cross-section of a self-interacting (scalar) particle is $\sigma = \lambda^2/16\pi m^2$. This should give a sufficient order of magnitude estimate for the cross-section for the vector particle, too. \\
For most calculations, we use the DD2 equation of state (with electrons) \cite{Hempel:2009mc}, taken from the CompOSE database \cite{Typel:2013rza}, to describe the NS component. It was chosen because it is widely used by a number of groups and thus is well known in the literature. The DD2 EOS is based on a relativistic mean-field model with density-dependent coupling constants, which has been fitted to the properties of nuclei and results from Brueckner-Hartree-Fock calculations for dense nuclear matter. Therefore, the DD2 EOS describes also the EOS of pure neutron matter from chiral effective field theory (see \cite{Kruger:2013kua}). For the purpose of our investigations, the particular choice of the nuclear equation of state is not of importance and has no effect on our general conclusions.

\begin{figure*}
\centering
    \includegraphics[width=0.495\textwidth]{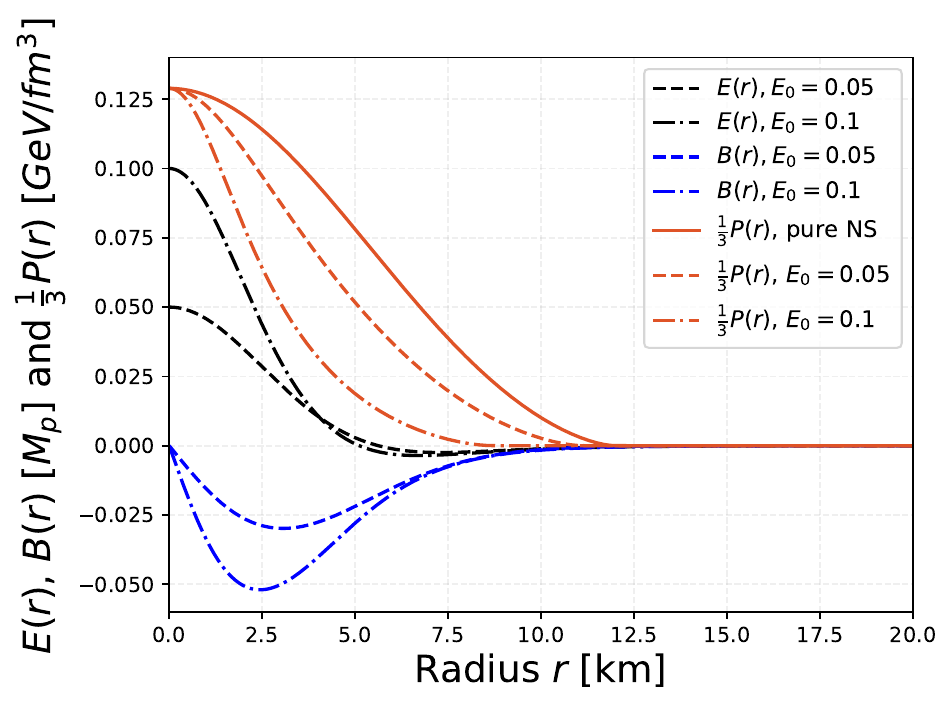}
    \includegraphics[width=0.495\textwidth]{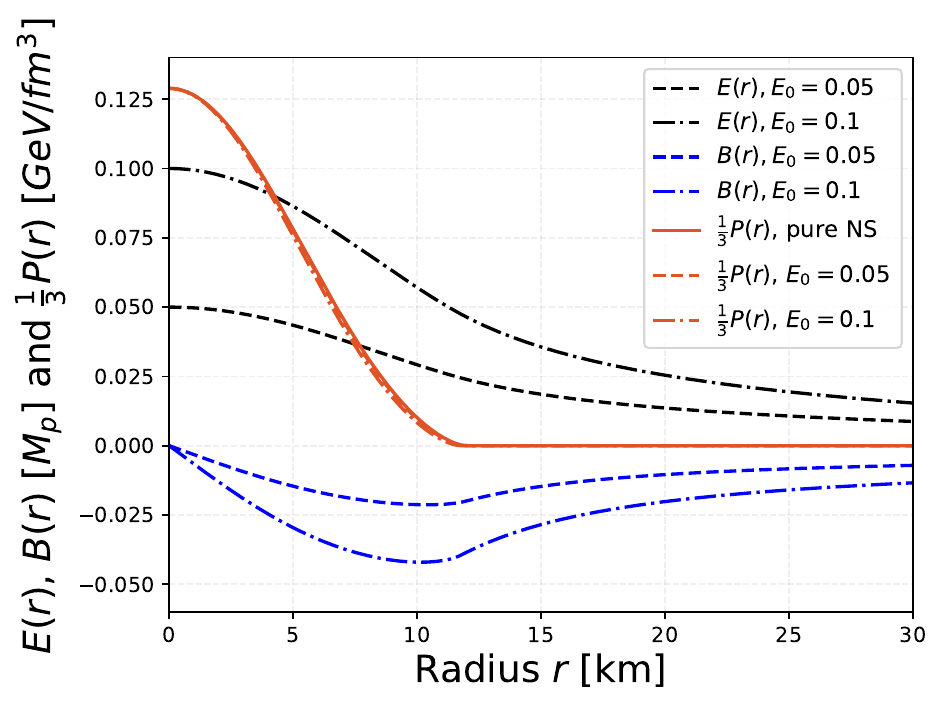}
    \caption{\textbf{Left panel:} Radial profiles of the pressure $P(r)$ (orange) and the vector field components $E(r)$ (black), $B(r)$ (blue) of the zeroth mode of different FPSs with potential \eqref{eq:fermion-boson-stars:fermion-proca-stars:quartic-self-interaction-potential}. The boson mass is $m=1.34 \e{-10}\,eV$ and $\Lambda_\mathrm{int} = 0$. The FPSs have a central density of $\rho_c = 5 \rho_{\mathrm{sat}}$ and varying central vector field amplitudes $E_0$. The pressure has been re-scaled by a factor of $3$ for convenience. The DM forms a core and compactifies the fermionic component.
    \textbf{Right panel:} Same as in the left panel, but this time the vector boson mass is set to $m=1.34 \e{-11}\,eV$. The DM forms a cloud around the fermionic component. The radius of the fermionic component is barely affected by the field. A kink can be seen in the profile for $B(r)$ at roughly $11.5\,km$. This corresponds to the point where the fermionic radius is located. This illustrates the gravitational back-reaction between the vector field and NS matter.}
    \label{fig:results:fermion-proca-stars:radial-profiles-1}
\end{figure*}

\begin{figure*}
\centering
    \includegraphics[width=0.495\textwidth]{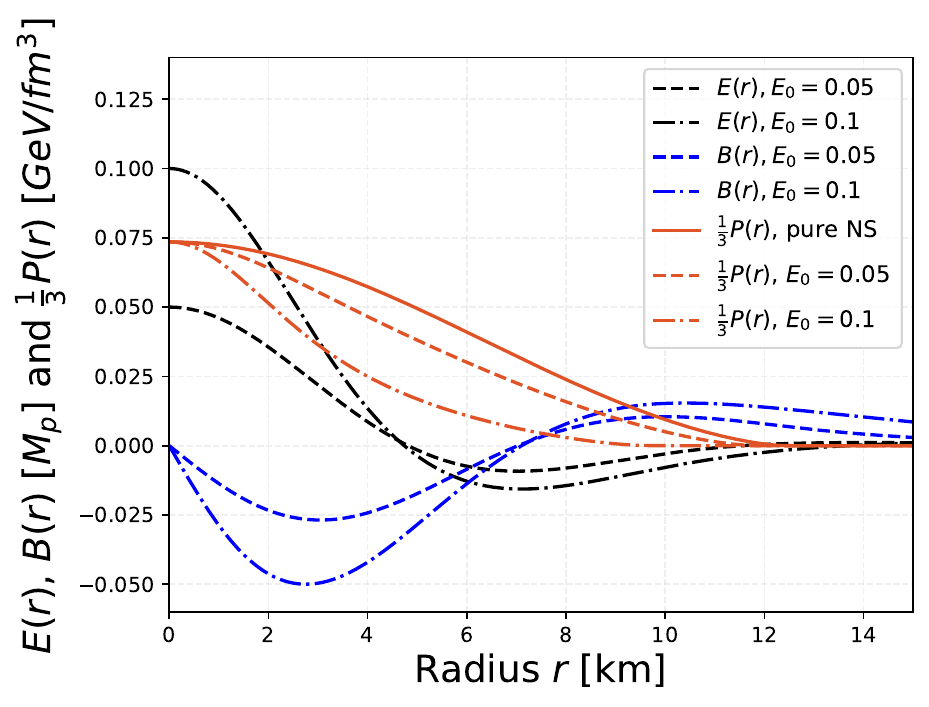}
    \includegraphics[width=0.495\textwidth]{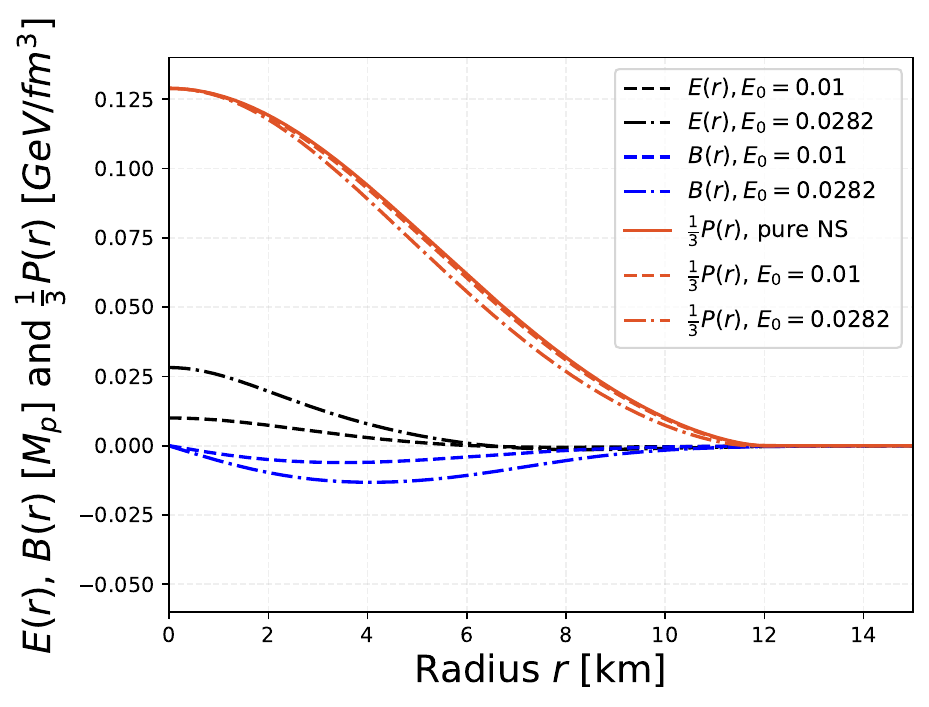}
    \caption{\textbf{Left panel:} Radial profiles of the pressure $P(r)$ (orange) and the vector field components $E(r)$ (black), $B(r)$ (blue) of the first mode of different FPSs with potential \eqref{eq:fermion-boson-stars:fermion-proca-stars:quartic-self-interaction-potential}. The boson mass is $m=1.005 \e{-10}\,eV$ and $\Lambda_\mathrm{int} = 0$. The FPSs have a central density of $\rho_c = 4 \rho_{\mathrm{sat}}$ and varying central vector field amplitudes $E_0$. The pressure has been re-scaled by a factor of $3$ for convenience.
    \textbf{Right panel:} Radial profiles of the pressure $P(r)$ (orange) and the vector field components $E(r)$ (black), $B(r)$ (blue) of FPSs in the zeroth mode with potential \eqref{eq:fermion-boson-stars:fermion-proca-stars:quartic-self-interaction-potential}. The boson mass is $m=1.34 \e{-10}\,eV$ and the self-interaction strength is $\Lambda_\mathrm{int} = 50$. The FPSs have a central density of $\rho_c = 5 \rho_{\mathrm{sat}}$ and varying central vector field amplitudes $E_0$. The pressure has been re-scaled by a factor of $3$ for convenience. Due to the analytical bound on $E_0$ \eqref{eq:fermion-boson-stars:fermion-proca-stars:analytical-bound-amplitude}, the maximal amplitude is roughly $E_{0,\text{crit}} \approx 0.0282$. The limited field amplitude strongly limits the effect on the fermionic component.}
    \label{fig:results:fermion-proca-stars:radial-profiles-2}
\end{figure*}

\subsection{Radial Profiles} \label{subsec:Radial_Profiles}

We compute the radial profiles of FPSs. In particular, we consider the radial dependence of the pressure $P(r)$ and the vector field components $E(r)$, $B(r)$. Even though the radial distribution of physical quantities can not yet be observed directly (although one could infer the DM distribution using the geodesic motion of light \cite{Shakeri:2022dwg}), a good understanding of the internal structure of FPSs can be used to deduce their global quantities and vice versa. Knowledge about the internal distribution is also relevant for numerical applications. Another reason we include the radial profiles here is to facilitate reproducibility of this work and for the sake of code-validation for future works. \\
Radial profiles of pure Proca stars have already been discussed by \cite{Brito:2015pxa} and for the case of a quartic self-interaction potential like \eqref{eq:fermion-boson-stars:fermion-proca-stars:quartic-self-interaction-potential} by \cite{Minamitsuji:2018kof}. We used the results of \cite{Minamitsuji:2018kof} in particular to verify that our code \cite{Diedrichs-Becker-Jockel} reproduces the results correctly and consistently. \\

In \autoref{fig:results:fermion-proca-stars:radial-profiles-1}, we show radial profiles of the pressure $P(r)$ (orange) and the vector field components $E(r)$ (black), $B(r)$ (blue) of the zeroth mode of different FPSs with potential \eqref{eq:fermion-boson-stars:fermion-proca-stars:quartic-self-interaction-potential}. In the left panel, we take a boson mass of $m=1.34 \e{-10}\,eV$ and an interaction strength of $\Lambda_\mathrm{int} = 0$. The FPSs have varying central vector field amplitudes $E_0$ and central densities of $\rho_c = 5 \rho_{\mathrm{sat}}$. Here we take $\rho_{\mathrm{sat}} = m_n n_{nuc} \approx 2.5\e{14}\,g/cm^3$ to be the nuclear saturation density, with $m_n$ being the neutron mass and $n_{nuc}=0.15\:$fm$^{-3}$ the average nuclear number density. The radial profile of a pure NS is shown with the orange continuous line and has no corresponding vector field (because it would be zero everywhere). The presence of the DM can be seen to compactify the NS component with increasing central field amplitude $E_0$. The field forms a DM core configuration. \\
In the right panel of \autoref{fig:results:fermion-proca-stars:radial-profiles-1}, all parameters are left equal except for the vector boson mass, which is set to $m=1.34 \e{-11}\,eV$. Due to the low DM mass, the correlation length increases, which increases the size of the vector field component and forms a DM cloud configuration. Since the amount of energy density of the vector field is distributed inside and outside the NS component, the effect on the radius is small. At around $r=11.5\,km$, a kink can be seen in the radial profile of the field component $B(r)$. This point coincides with the point where the fermionic radius of the FBS is located. This illustrates the gravitational back-reaction between the vector field and the NS component of the FBS. \\

In \autoref{fig:results:fermion-proca-stars:radial-profiles-2}, we show radial profiles of the pressure $P(r)$ (orange) and the vector field components $E(r)$ (black), $B(r)$ (blue) of an FPS. In the left panel, we show an FPS in the first mode, which can be identified by the fact that the $E(r)$ component crosses the x-axis twice and $B(r)$ crosses it once. The boson mass is $m=1.005 \e{-10}\,eV$ and $\Lambda_\mathrm{int} = 0$. This time, the central density is taken to be $\rho_c = 4 \rho_{\mathrm{sat}}$ and the central vector field amplitudes vary. \\
The right panel of \autoref{fig:results:fermion-proca-stars:radial-profiles-2} shows an FPS in the zeroth mode with a vector boson mass of $m=1.34 \e{-10}\,eV$ and a self-interaction strength of $\Lambda_\mathrm{int} = 50$. The maximal amplitude is roughly $E_{0,\text{crit}} \approx 0.0282$ due to the analytical bound on $E_0$, see \eqref{eq:fermion-boson-stars:fermion-proca-stars:analytical-bound-amplitude}. The limited field amplitude strongly limits the possible effect on the fermionic component and thus on the fermionic radius, especially in the limit of large $\Lambda_\mathrm{int}$. It may therefore be difficult to detect strongly self-interacting vector DM within a NS if one only considers measurements of the fermionic radius. It is also conceivable that the maximum amplitude $E_{0,\text{crit}}$ implies a maximum amount of possible accretion of vector DM, which could be used to set bounds on the DM self-interaction strength. We leave the analysis of this aspect for a future work.

\subsection{Stable Solutions} \label{subsec:Stable_Solutions}

\begin{figure*}
\centering
    \includegraphics[width=0.493\textwidth]{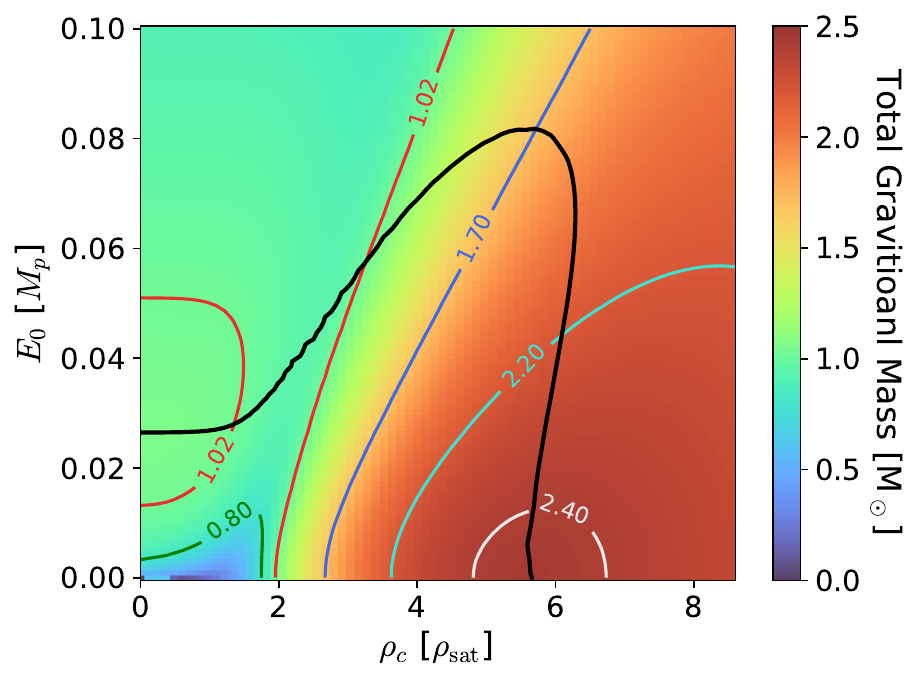}
    \includegraphics[width=0.48\textwidth]{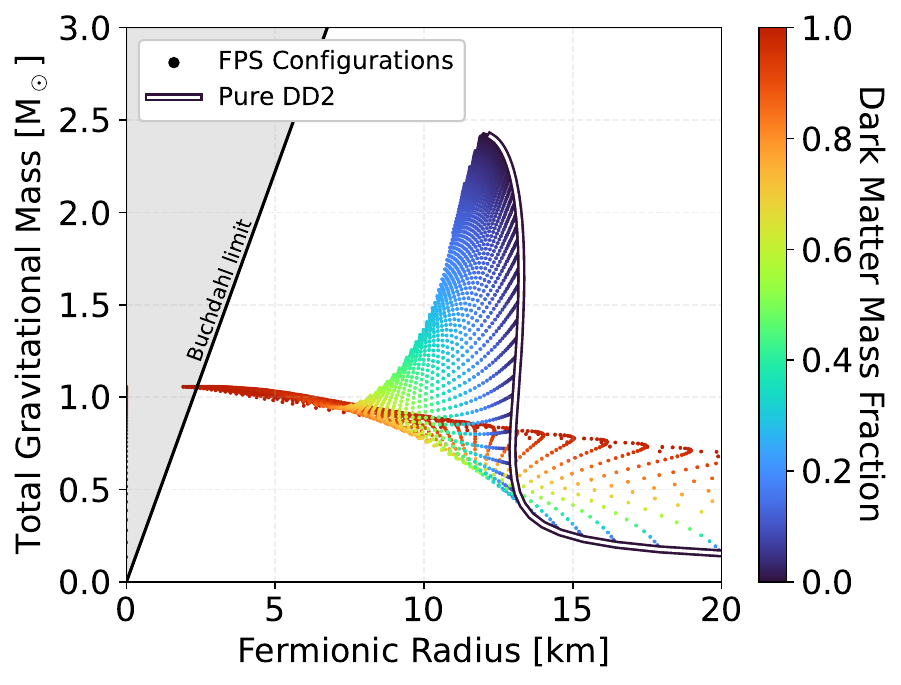}
    \caption{\textbf{Left panel:} Total gravitational mass of different FPSs as a function of the restmass density $\rho_c$ and central vector field amplitude $E_0$. Additionally we show contours of constant gravitational mass. The black line corresponds to the stability curve, which separates stable solutions (in the lower left region) from unstable solutions (everywhere else).
    \textbf{Right panel:} Mass-radius diagram displaying the fermionic radius vs the total gravitational mass for FPS configurations that are within the stability region displayed in the left panel. Each point corresponds to a single configuration and is colour-coded according to the DM-fraction $N_\mathrm{b}/(N_\mathrm{b} + N_\mathrm{f})$. The solid black-white line shows the mass-radius curve for pure fermionic matter. In both cases, a vector field with mass of $m=1.34 \e{-10}\,eV$ and no self-interactions was considered in addition to the DD2 EOS for the fermionic part.}
    \label{fig:results:fermion-proca-stars:stability-and-MR-curve-lamda0}
\end{figure*}

\begin{figure*}
\centering
    \includegraphics[width=0.493\textwidth]{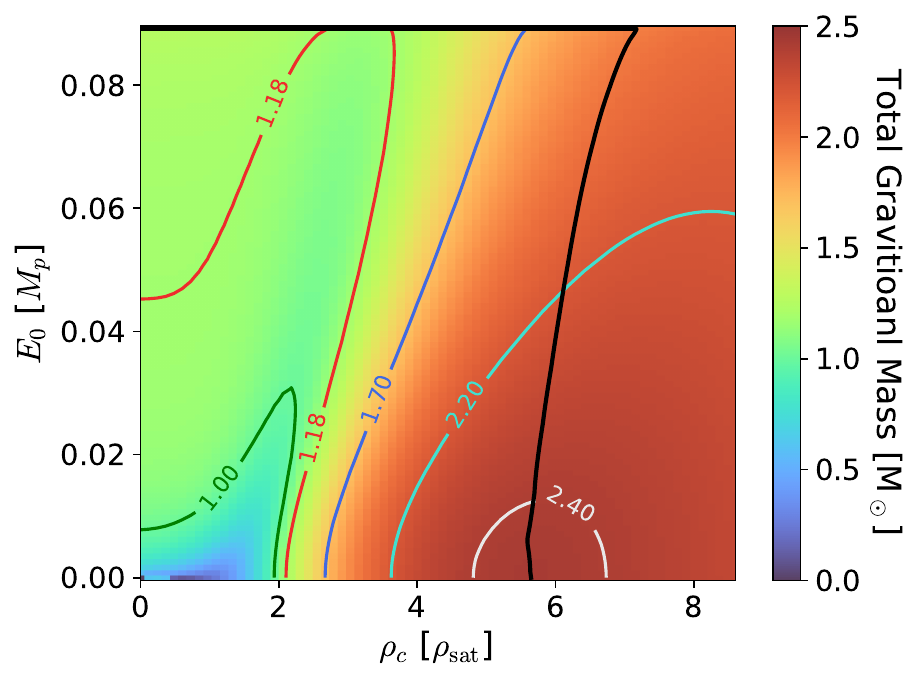}
    \includegraphics[width=0.48\textwidth]{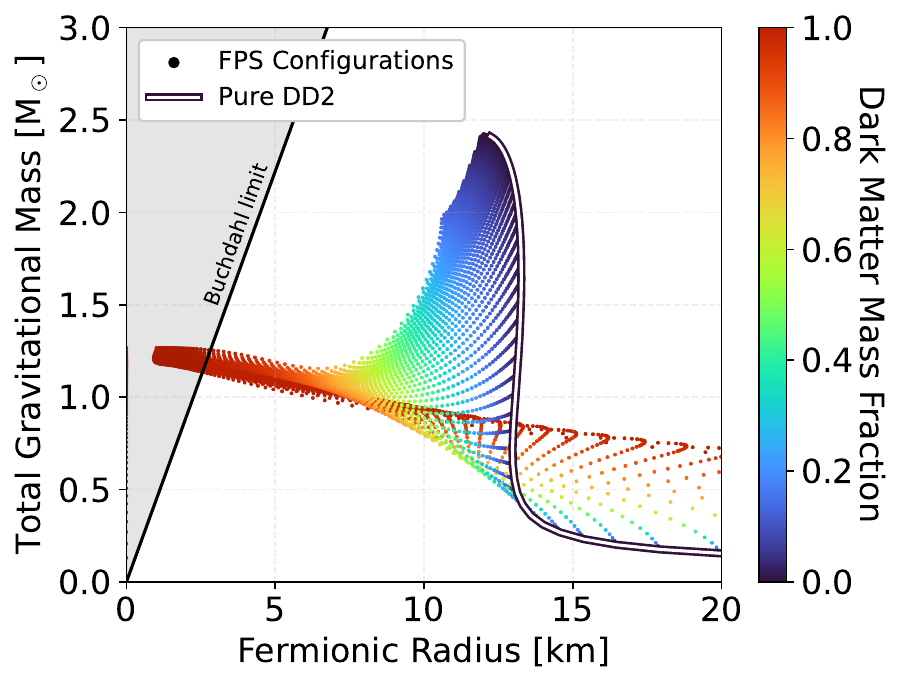}
    \caption{\textbf{Left panel:} Total gravitational mass of different FPSs as a function of the restmass density $\rho_c$ and central vector field amplitude $E_0$, with $m=1.34 \e{-10}\,eV$ and $\Lambda_\mathrm{int}=5$. Additionally we show contours of constant gravitational mass. The black line corresponds to the stability curve, which separates stable solutions (in the lower left region) from unstable solutions (everywhere else). The stability curve reaches configurations with the maximum possible vector field amplitude $E_{0,\text{crit}} \approx 0.089$. This is a feature unique to FPSs.
    \textbf{Right panel:} Mass-radius diagram displaying the fermionic radius vs the total gravitational mass for FPS configurations that are within the stability region displayed in the left panel. Each point corresponds to a single configuration and is colour-coded according to the DM-fraction $N_\mathrm{b}/(N_\mathrm{b} + N_\mathrm{f})$. The solid black-white line shows the mass-radius curve for pure fermionic matter. A vector field with mass of $m=1.34 \e{-10}\,eV$ and $\Lambda_\mathrm{int}=5$ was considered in addition to the DD2 EOS for the fermionic part.}
    \label{fig:results:fermion-proca-stars:stability-and-MR-curve-lamda5}
\end{figure*}

We compute a grid of FPSs with different central densities $\rho_c$ and central vector field amplitudes $E_0$. Using the array of solutions, we compute the stability curve using the stability criterion \eqref{eq:fermion-boson-stars:scalar-fermion-boson-stars:FBS-stability-criterion-rewritten}. The stable solutions can then be filtered and analyzed further. \\
This can be seen in the left panel of \autoref{fig:results:fermion-proca-stars:stability-and-MR-curve-lamda0}, where we compute FPSs with a quartic self-interaction potential \eqref{eq:fermion-boson-stars:fermion-proca-stars:quartic-self-interaction-potential} with $m=1.34 \e{-10}\,eV$ and $\Lambda_\mathrm{int}=0$. We additionally compute the stability curve using the stability criterion \eqref{eq:fermion-boson-stars:scalar-fermion-boson-stars:FBS-stability-criterion-rewritten}. The stability curve defines the boundary between stable and unstable configurations under linear radial perturbations. The shape of the stability curve for FPSs is qualitatively very similar to the case of scalar FBSs (compare to \cite{Diedrichs:2023trk}). For pure neutron stars and Proca stars, respectively, the curve converges on the $\rho_c$- and $E_0$-axis at the point, where the non-mixed configurations have their maximum gravitational masses. We take only the FPSs inside the stability region, enclosed by the stability curve, and plot them in a mass-radius (MR) diagram. This leads to the graph in the right panel of \autoref{fig:results:fermion-proca-stars:stability-and-MR-curve-lamda0}. \\
We see that stable FPS configurations form an MR region instead of an MR curve (which would be the case for single-fluid systems). The stable configurations form core or cloud solutions, depending on their DM-fraction $N_\mathrm{b}/(N_\mathrm{b} + N_\mathrm{f})$. A DM core is present if the bosonic component is geometrically smaller than the fermionic component (i.e. when $R_\mathrm{b} < R_\mathrm{f}$). The opposite is true for cloud configurations. The FPSs with high DM-fractions have masses of roughly $1\,M_\odot$. This is higher than for scalar FBSs with equal boson mass $m$ (compare to \cite{Diedrichs:2023trk}). This can be explained through the different scaling relations for pure Proca stars and boson stars. \\

Another point where FPSs differ from FBSs is the existence of a maximal amplitude $E_{0,\text{crit}}$ \eqref{eq:fermion-boson-stars:fermion-proca-stars:analytical-bound-amplitude} for the vector field. When increasing the self-interaction strength $\Lambda_\mathrm{int}$, the maximal possible vector field amplitude shrinks. This affects the shape of the stability curve. \\
In \autoref{fig:results:fermion-proca-stars:stability-and-MR-curve-lamda5} (left panel), we show such a case where the self-interaction strength is $\Lambda_\mathrm{int} = 5$. The stability curve does not reach the $E_0$-axis anymore, but instead rises vertically from the pure NS configurations until it reaches the FPSs with maximal central vector field amplitude $E_{0,\text{crit}} \approx 0.089$. We have manually extended the stability curve so that it proceeds horizontally until it reaches the $E_0$-axis. It is noteworthy that this behavior starts at surprisingly small self-interaction strengths and persists up to higher $\Lambda_\mathrm{int}$. \\
In principle, also a third behavior of the stability curve of FPSs is conceivable. For some specific $\Lambda_\mathrm{int}$, it should be possible that the stability curve does not admit one continuous shape like in \autoref{fig:results:fermion-proca-stars:stability-and-MR-curve-lamda0} or \autoref{fig:results:fermion-proca-stars:stability-and-MR-curve-lamda5}, but that the stability curve is cut into two parts. Namely, one part which starts at the $E_0$-axis and then rises to reach the edge where $E_{0,\text{crit}}$ is located, and another part which starts at the $\rho_c$-axis and then rises roughly vertically until it too reaches the analytical bound for the vector field amplitude $E_{0,\text{crit}}$ (think of a horizontal line cutting through the stability curve in \autoref{fig:results:fermion-proca-stars:stability-and-MR-curve-lamda0} at, e.g., $E_0 = 0.06$). During our testing, we did not find any case where the stability curve follows this behavior. However, there is also no reason that we are aware of why such a behavior of the stability curve should be forbidden. This is why we presume that such a case might exist. \\

\begin{figure*}
    \centering
    \includegraphics[width = 0.995\textwidth]{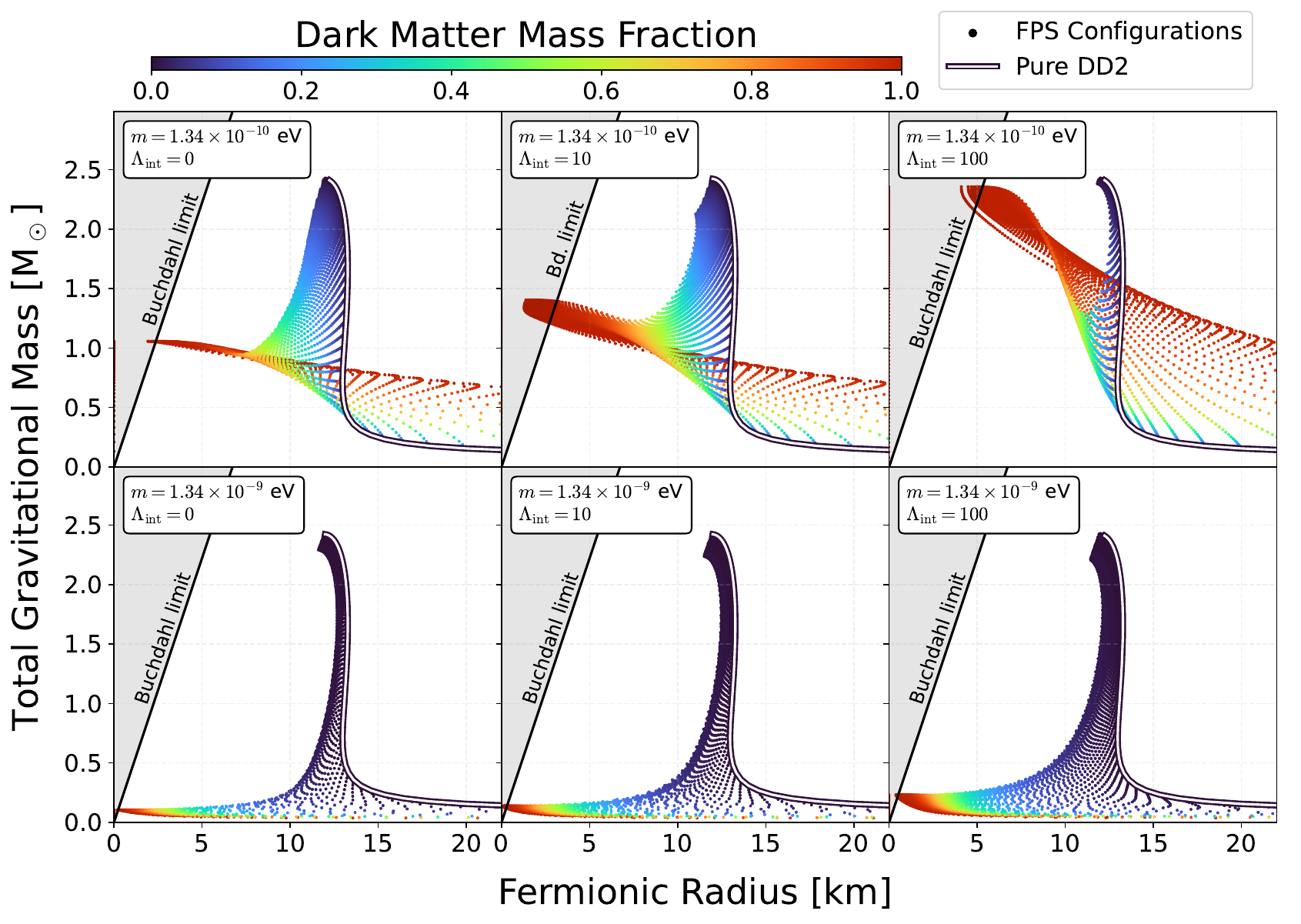}
    \includegraphics[width = 0.995\textwidth]{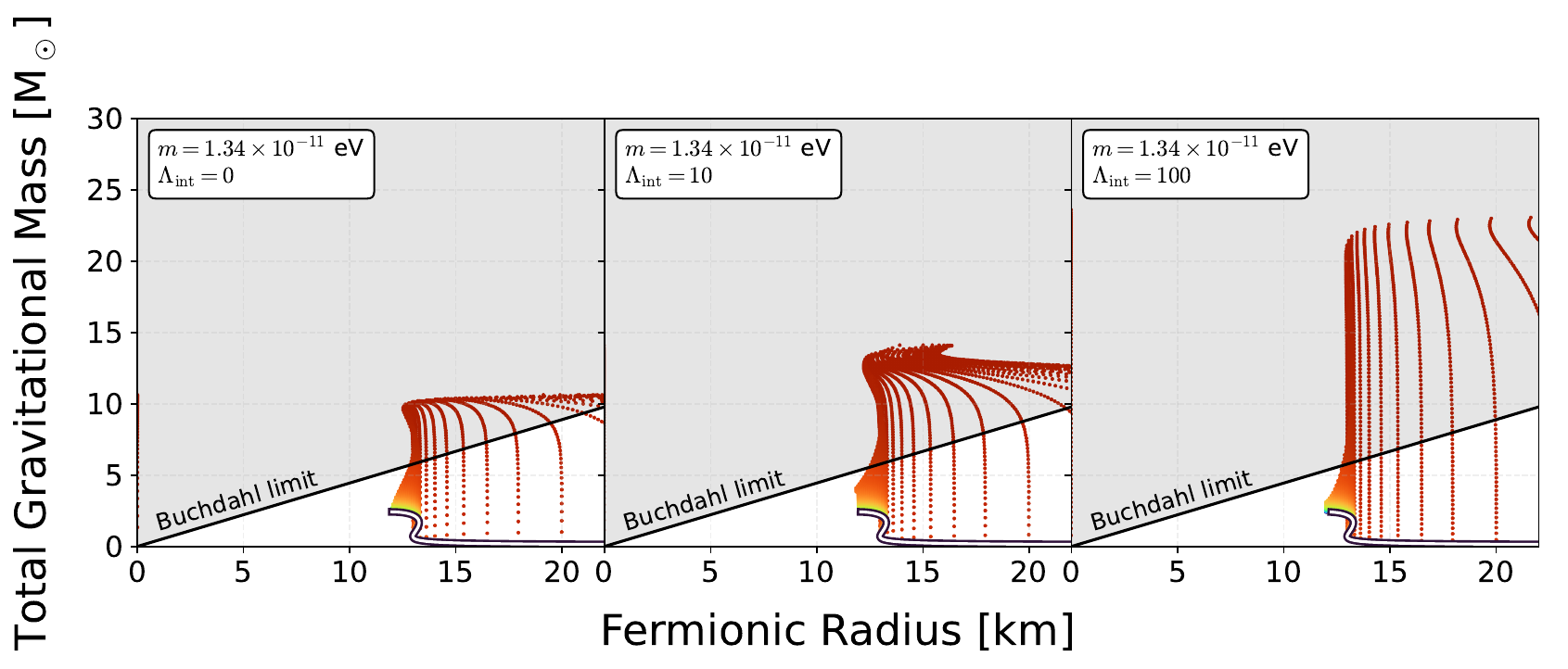}
    \caption{Relation between total gravitational mass $M_\mathrm{tot}$ and fermionic radius $R_\mathrm{f}$ for different FPSs. The rows correspond to bosonic masses of $m = \{1, 10, 0.1\}\times 1.34 \e{-10}\,eV$, columns correspond to self-interactions of $\Lambda_\mathrm{int}= \{0, 10, 100\}$ respectively. We use the DD2 EOS for the fermionic part. Notice the different scale of the bottom plots. The gray region marks the Buchdahl limit, where no stable NS can exist. Observing only $R_\mathrm{f}$ of these systems would appear to violate the Buchdahl limit, even though the FPS as a whole does not.}
    \label{fig:results:fermion-proca-stars:MR-grid}
\end{figure*}

\begin{figure*}
    \centering
    \includegraphics[width = 0.995\textwidth]{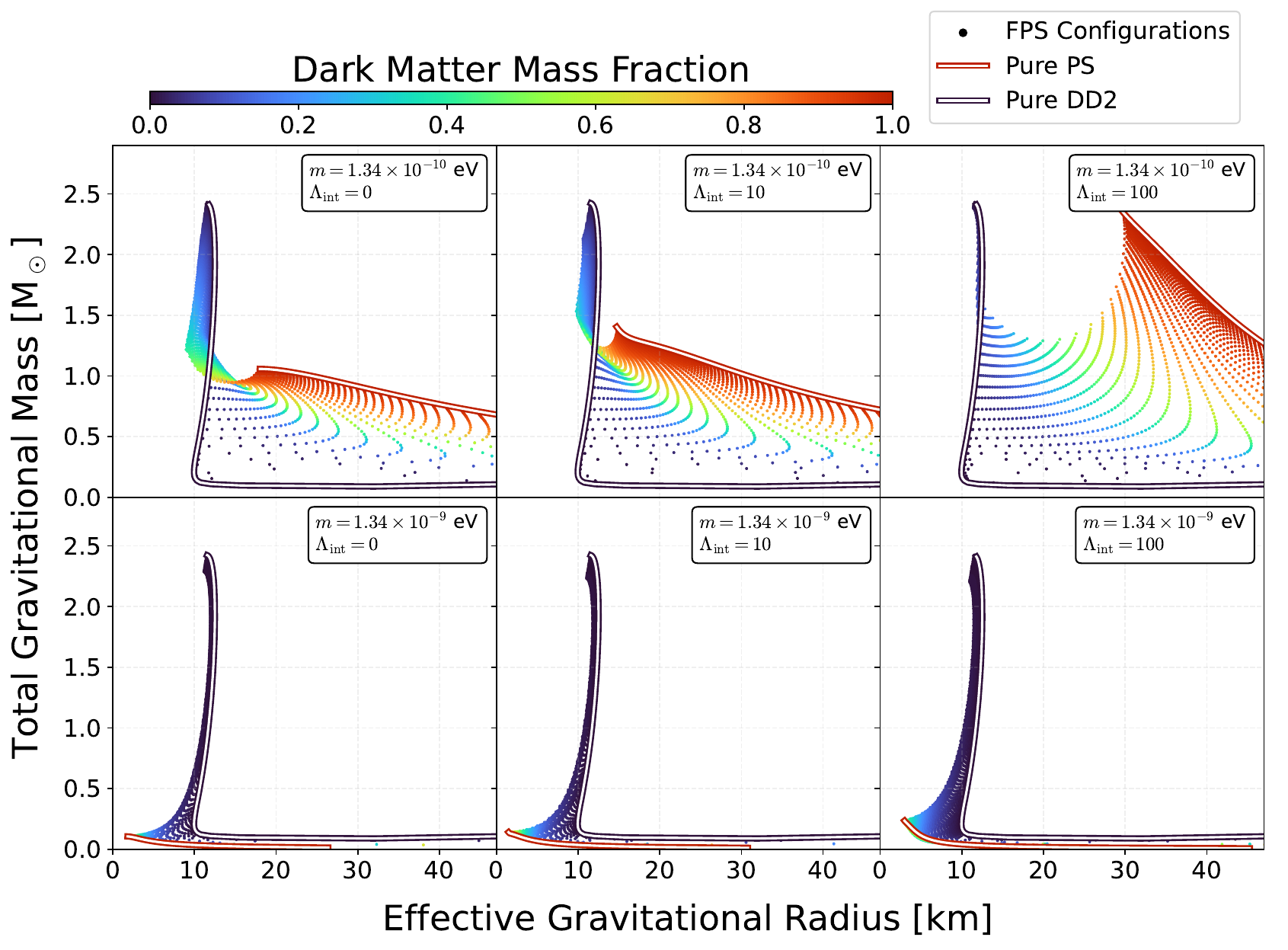}
    \includegraphics[width = 0.995\textwidth]{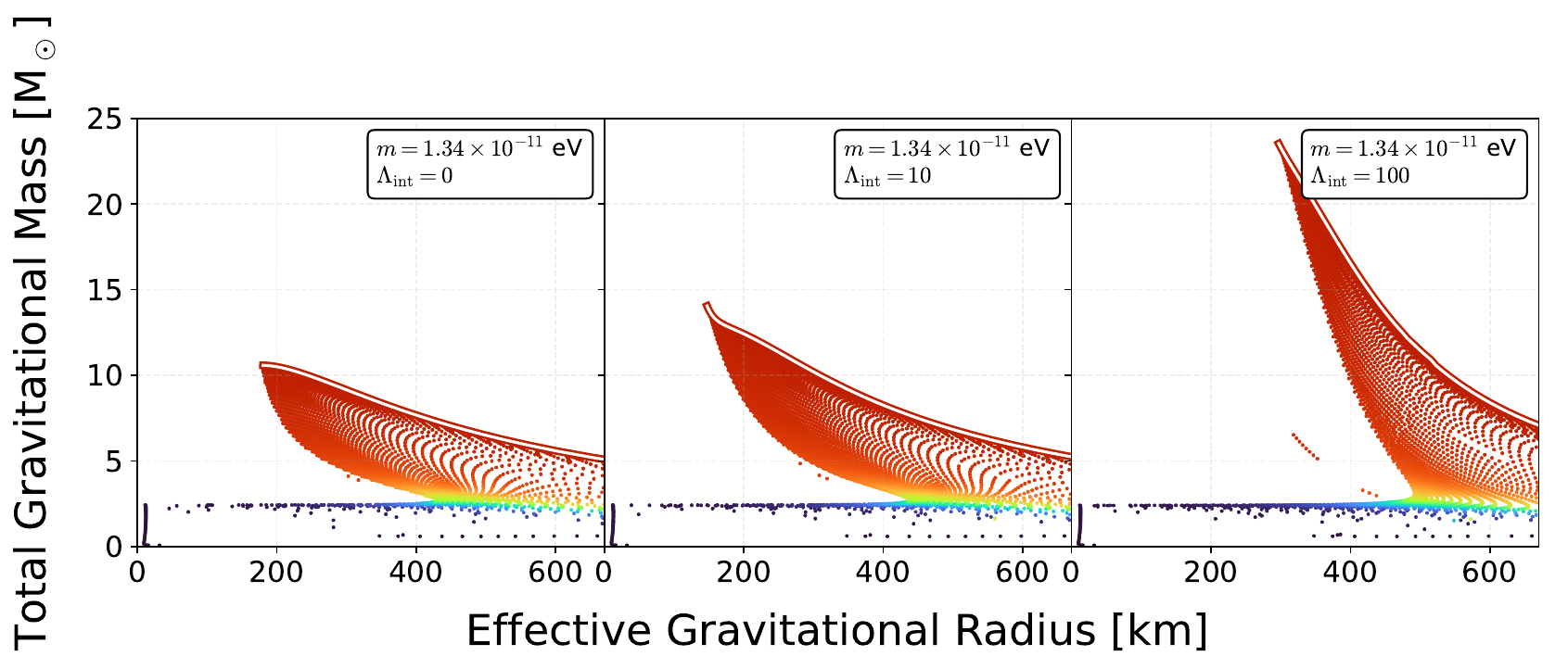}
    \caption{Relation between total gravitational mass $M_\mathrm{tot}$ and effective gravitational radius $R_G$ for different FPSs. $R_G$ is the radius where $99\%$ of the total rest mass is contained. The rows correspond to bosonic masses of $m = \{1, 10, 0.1\}\times 1.34 \e{-10}\,eV$, columns correspond to self-interactions of $\Lambda_\mathrm{int}= \{0, 10, 100\}$ respectively. We use the DD2 EOS for the fermionic part. Notice the different scales of the bottom plots. For pure NSs, because the crust has comparatively low density, $R_G$ is significantly smaller than $R_\mathrm{f}$ (compare to \autoref{fig:results:fermion-proca-stars:MR-grid}). $R_G$ tends to be higher as compared to scalar FBSs for equal boson masses and self-interaction strength (compare to Figure 3 in \cite{Diedrichs:2023trk}).}
    \label{fig:results:fermion-proca-stars:MRg-grid}
\end{figure*}

We compute various FPSs with different values of the vector boson masses $m = \{1, 10, 0.1\}\times 1.34 \e{-10}\,eV$ and self-interaction strengths $\Lambda_\mathrm{int}= \{0, 10, 100\}$. We chose the same parameter values as in \cite{Diedrichs:2023trk} to allow for easy comparability. In \autoref{fig:results:fermion-proca-stars:MR-grid}, we show the mass and fermionic radii of all stable FPS configurations in an MR diagram. In \autoref{fig:results:fermion-proca-stars:MRg-grid}, we show the mass plotted against the effective gravitational radius $R_G$. It is defined as the radius where $99\,\%$ of the total rest mass $N_\mathrm{f}+N_\mathrm{b}$ is contained. The stable solutions have been obtained using the stability criterion \eqref{eq:fermion-boson-stars:scalar-fermion-boson-stars:FBS-stability-criterion-rewritten}. Note the different axis scaling in the figures. It was chosen such that the relevant trends and features of the solutions can be seen well. \\
We hereafter discuss some general trends and compare the results to the one obtained for scalar FBSs. The following analysis should thus be explicitly compared to figures 2 and 3 in \cite{Diedrichs:2023trk}. \\
We find that many of the general conclusions regarding FBSs can also be applied to FPSs. FPSs with small DM-fractions are dominated by the fermionic component, leading to only small changes in the fermionic radius. In the case of DM dominated FPSs, the solutions behave similar to pure Proca stars. This leads to higher masses as compared to FBSs, where the total gravitational mass of pure boson stars will be roughly half that of a Proca star with the same boson mass, as can be seen well for the cases where $m = \{1, 0.1\}\times 1.34 \e{-10}\,eV$. FPSs can thus reach higher total gravitational masses as compared to FBSs with the same DM mass and self-interaction strength. \\
For $m = 1.34 \e{-9}\,eV$, the bosonic component is concentrated inside the fermionic one and forms a DM core. Even small amounts of DM can have a significant impact on the fermionic radius, since the whole vector field is concentrated entirely inside the NS component. More massive DM particles can thus have larger effects on the fermionic radius compared to low-mass DM at similar DM-fractions. This is due to the cloud-like structure of low-mass DM. For small DM  masses, the majority of the DM will be concentrated outside the NS part -- due to its larger correlation length -- and will thus have smaller effects on the fermionic radius. The smaller the mass and the larger the self-interaction strength, the more likely the formation of a DM cloud is. The opposite is true for DM core solutions. FPSs tend to produce configurations with larger total masses compared to scalar FBSs. Their halos also extend to larger radii, as can be seen from the gravitational radius in \autoref{fig:results:fermion-proca-stars:MRg-grid}. \\
In general, the gravitational radius of FPSs is larger in size as compared to scalar FBSs (compare to Figure 3 in \cite{Diedrichs:2023trk}). The larger gravitational radius suggests that FPS have larger tidal deformabilities, compared to their scalar field counterparts (FBS) with equal $m$ and $\Lambda_\mathrm{int}$. This is because objects with larger radii are generally favored to tidally disrupt. This could favor higher vector boson masses compared to the corresponding scalar boson mass in the case of FBSs. A future quantitative analysis of the tidal deformability of FPSs is needed to definitively verify this hypothesis. \\
When considering the gravitational radius of FPSs with small boson masses of $m = 1.34 \e{-11}\,eV$ (bottom row of \autoref{fig:results:fermion-proca-stars:MRg-grid}), the transition between DM-dominated and NS-dominated configurations appears more abrupt than in the FBS case (compare to Figure 3 in \cite{Diedrichs:2023trk}). For example, when starting with a system with a DM-fraction of roughly $0\%$ or $80\%$, increasing the DM-fraction by small amounts can massively impact the total mass and gravitational radius of the combined system. \\
Finally, note the outlier points in \autoref{fig:results:fermion-proca-stars:MRg-grid} for $m = 1.34 \e{-11}\,eV$ and $\Lambda_\mathrm{int}=100$ at roughly $R_G=350\,km$. These are likely to be numerical artifacts and should thus not be regarded as physical. This is to be expected since for small DM masses and large self-interactions, the numerical solution gets increasingly difficult. This problem could be avoided by using smaller step-sizes and higher numerical precision. But this would also lead to longer run-times of the code.

\subsection{Comparison with Scalar FBS} \label{subsec:Comparison_Scalar_FBS}

\begin{figure*}
\centering
    \includegraphics[width=0.495\textwidth]{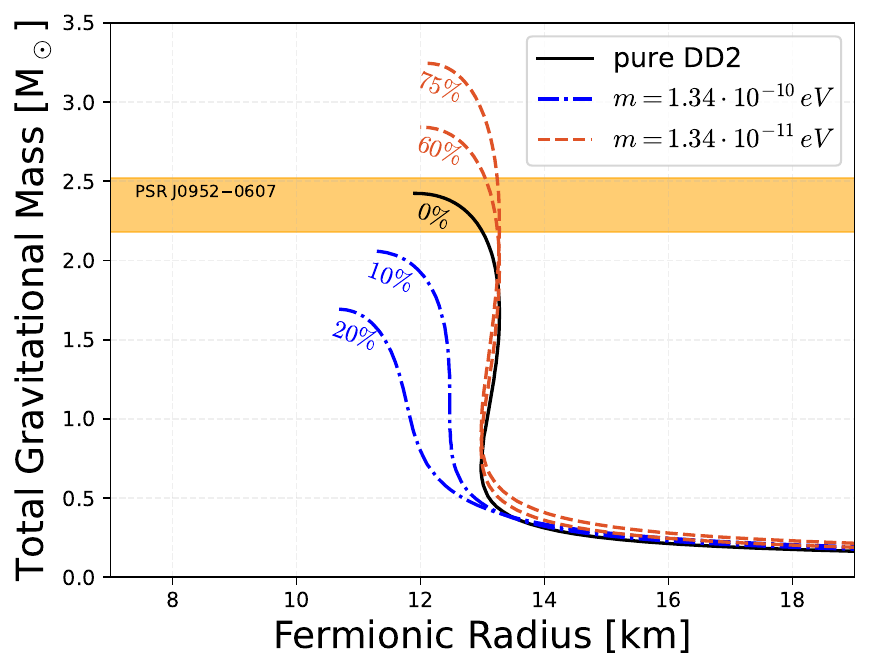}
    \includegraphics[width=0.495\textwidth]{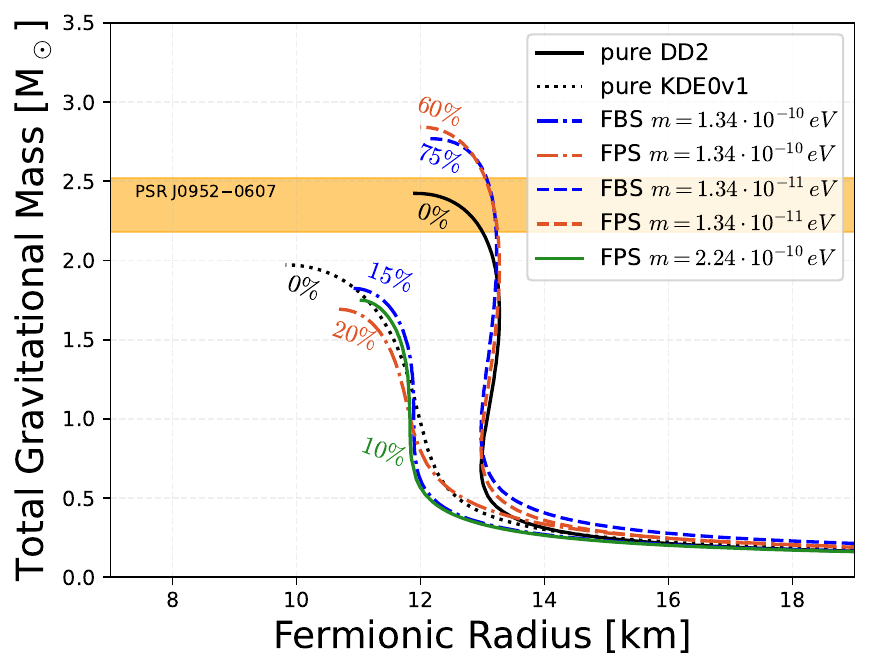}
    \caption{\textbf{Left panel:} Mass-radius relations of FPSs with the DD2 EOS \cite{Hempel:2009mc} for vector boson masses $m = \{1, 0.1\}\times 1.34 \e{-10}\,eV$, no self-interactions and constant DM-fractions $N_\mathrm{b}/(N_\mathrm{b} + N_\mathrm{f})$. This figure should be compared to Figure 5 (left panel) in \cite{Diedrichs:2023trk} as the same masses and DM-fractions were chosen. The orange band marks the observational constraint of J0952-0607 \cite{Romani:2022jhd} and the percentage numbers denote the respective DM-fractions.
    \textbf{Right panel:} Mass-radius relations of FPSs (orange and green lines) and FBSs (blue lines) with the DD2 EOS for different boson masses, no self-interactions and different DM-fractions. The black lines correspond to the pure NSs with the DD2 EOS and KDE0v1 EOS \cite{Schneider:2017tfi} respectively. FPS and FBS solutions with different masses and DM-fractions can both be degenerate with each other, or also degenerate with pure NSs with a different EOS.}
    \label{fig:results:fermion-proca-stars:MR-curves-Nbfrac}
\end{figure*}

\begin{figure*}
\centering
    \includegraphics[width=0.495\textwidth]{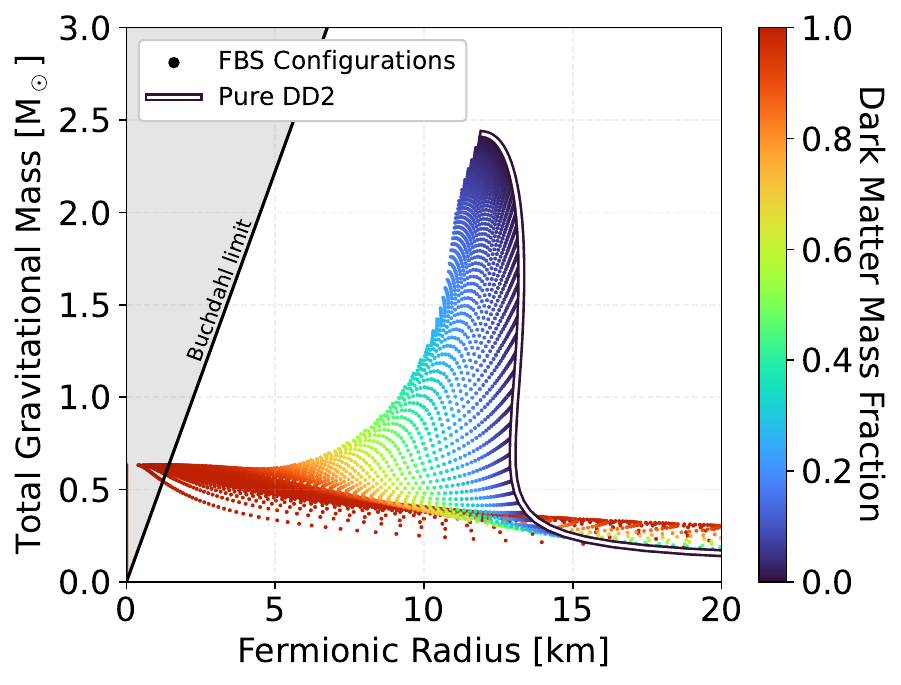}
    \includegraphics[width=0.495\textwidth]{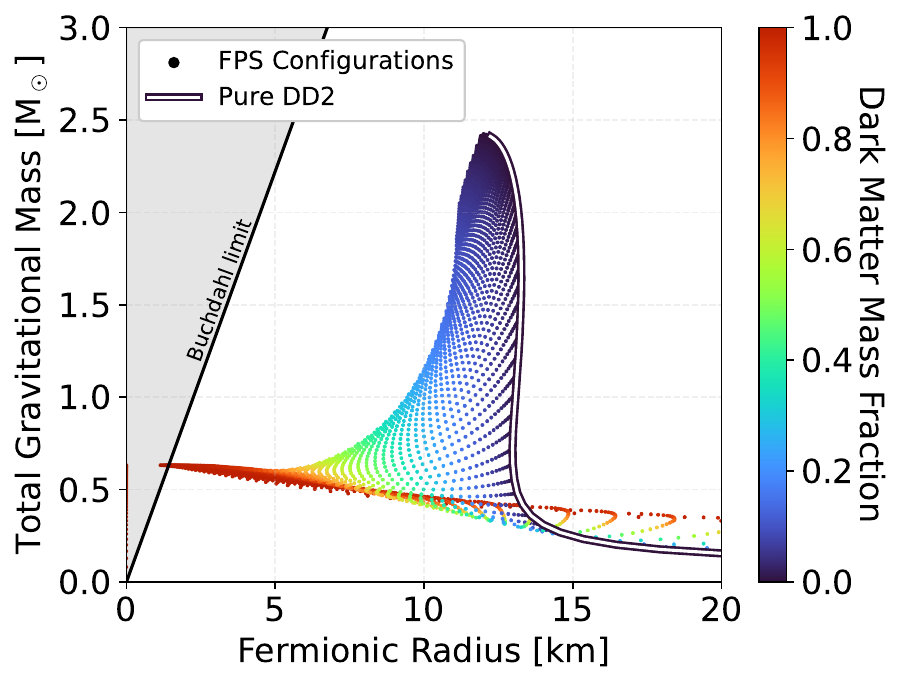}
    \caption{\textbf{Left panel:} Mass-radius diagram displaying the fermionic radius vs. the total gravitational mass for stable FBS configurations with scalar boson mass of $m = 1.34 \e{-10}\,eV$ and no self-interaction. Each point corresponds to a single configuration and is color-coded according to the DM-fraction $N_\mathrm{b}/(N_\mathrm{b} + N_\mathrm{f})$. The solid black-white line shows the mass-radius curve for pure fermionic matter, modeled by the DD2 EOS.
    \textbf{Right panel:} Mass-radius diagram displaying the fermionic radius vs. the total gravitational mass for stable FPS configurations with vector boson mass of $m = 2.24 \e{-10}\,eV$ and no self-interaction. Each point corresponds to a single configuration and is color-coded according to the DM-fraction $N_\mathrm{b}/(N_\mathrm{b} + N_\mathrm{f})$. The solid black-white line shows the mass-radius curve for pure fermionic matter. The vector boson mass was chosen so that in the limit of pure boson stars/Proca stars, the same total gravitational mass is produced. Both diagrams show only marginal differences.}
    \label{fig:results:fermion-proca-stars:MR-curves-FBS-FPS-comparison}
\end{figure*}

We show MR relations of FPSs and scalar FBSs with fixed DM-fractions $N_\mathrm{b}/(N_\mathrm{b} + N_\mathrm{f})$. \\
In the left panel of \autoref{fig:results:fermion-proca-stars:MR-curves-Nbfrac}, we show different FPSs with constant DM-fractions. The DD2 EOS \cite{Hempel:2009mc} was used for the NS component. For the vector boson, we chose masses of $m = \{1, 0.1\}\times 1.34 \e{-10}\,eV$ and no self-interactions. This figure should be explicitly compared to Figure 5 (left panel) in \cite{Diedrichs:2023trk} as the same masses and DM-fractions were chosen. The MR curve of a pure NS with the DD2 EOS (black line) is shown as a reference. Depending on the boson mass, FPSs can have increased or decreased maximum total gravitational mass when there is vector DM present. FPSs tend to produce configurations with larger gravitational masses compared to FBSs with equal parameters (mass, self-interaction and DM-fraction). This is not surprising when considering the scaling relations of pure boson stars and Proca stars, respectively. The gravitational mass scales like $M_\mathrm{max} \approx 0.633 M^2_p / m$ for pure boson stars and like $M_\mathrm{max} \approx  1.058 M^2_p / m$ for pure Proca stars, where $m$ is the mass of the scalar/vector boson, respectively. The presence of light bosonic DM can help to increase the total gravitational mass of a NS. This can make EOS which do not fulfill the observational constraints for the maximum NS mass viable again. Vector DM has a larger effect on the gravitational mass than scalar DM and thus smaller amounts of vector DM are needed to produce an equal increase in the total gravitational mass. \\
In the right panel of \autoref{fig:results:fermion-proca-stars:MR-curves-Nbfrac}, we show different FPSs (orange and green lines) and FBSs (blue lines) for different boson masses, no self-interactions and constant DM-fractions $N_\mathrm{b}/(N_\mathrm{b} + N_\mathrm{f})$. We used the DD2 EOS \cite{Hempel:2009mc} for the NS component. The parameters were chosen in a way to illustrate the degeneracies that can arise from different DM models or EOS for the NS component. For example, FPSs and FBSs with boson masses of $m = 1.34 \e{-11}\,eV$ (dashed lines) produce virtually indistinguishable mass-radius relations, when the FPSs and the FBSs have a DM-fraction of $60\%$ and $75\%$ respectively. A similar behavior can be seen for the cases where the boson mass is $m = 1.34 \e{-10}\,eV$ (dot-dashed lines). Here, FBSs with $15\%$ DM-fraction produce similar MR curves to FPSs with $20\%$ DM-fraction. In addition, the resulting MR curves are comparable to the curve corresponding to a pure NS with the KDE0v1 EOS \cite{Schneider:2017tfi}. They also match the curve corresponding to an FPS with $10\%$ DM-fraction and a vector boson mass of $m = 2.24 \e{-10}\,eV$ (green line). \\
In conclusion, FPSs can produce degenerate results in the MR plane with both FBSs and pure NS, given that different DM-fractions and EOS are allowed. Additional observables, such as the tidal deformability, are needed to break the degeneracy. However, it seems difficult to prevent degenerate solutions from existing in general, since FPSs themselves can be degenerate with other FPS-solutions that have different boson masses and DM-fractions. \\

We further explore the degeneracy between FPS and FBS solutions. In \autoref{fig:results:fermion-proca-stars:MR-curves-FBS-FPS-comparison}, we show the stable FBS and FPS solutions in an MR diagram. We used the scaling relations of the maximum mass for pure boson stars ($M_\mathrm{max} \approx 0.633 M^2_p / m$) and pure Proca stars ($M_\mathrm{max} \approx  1.058 M^2_p / m$) to match the boson masses in a way that both FPSs and FBSs will have the same gravitational mass in the pure boson star/Proca star limit. To guarantee matching solutions in this limit, we chose a scalar boson mass of $m = 1.34 \e{-10}\,eV$ and we chose a mass of $1.058 \div 0.633 \approx 1.671$ times the mass of the scalar boson -- i.e. $m = 2.24 \e{-10}\,eV$ -- for the vector boson. We find a high degree of similarity between the MR region of FBSs and FPSs with the scaled masses. This makes both solutions almost indistinguishable. The small differences present between the left and right panel of \autoref{fig:results:fermion-proca-stars:MR-curves-FBS-FPS-comparison} can be attributed to a slightly different grid-spacing used for the initial conditions $\rho_c$, $\phi_c$ (and $\rho_c$, $E_0$). This can be seen in the MR regions at small total gravitational masses $M_\mathrm{tot} <0.5\,M_\odot$ and also at radii $R_\mathrm{f} > 15\,km$. The color shading further reveals a different distribution of DM-fractions for a given $M_\mathrm{tot}$ and $R_\mathrm{f}$, even though the difference is small. \\
We expect a similar behavior to hold when considering different scalar and vector boson masses (with zero-self-interaction), given that they differ by the same factor of $\approx 1.671$. This adds further confidence to the observation that FBSs and FPSs might be difficult to distinguish since a given solution might be another system but with different boson mass (or DM-fraction). \\
Similar scaling relations also exist for boson stars and Proca stars in the limit of large self-interactions $\Lambda_\mathrm{int}$. A similar procedure might therefore be possible when also matching the self-interaction strength appropriately. An independent measurement of the DM particle mass would break this degeneracy to a certain degree. But it would also be necessary to constrain the self-interaction strength and the DM-fraction through other means. For example using correlations of the DM abundance in the galactic disk (see \cite{Giangrandi:2022wht,Sagun:2022ezx}) or using the bound on the maximal vector field amplitude. \\
The scaling behavior between (fermion) boson stars and (fermion) Proca stars also suggests another application. If it persists for large non-zero self-interactions, it might be possible to use the effective bosonic EOS derived by Colpi et al. \cite{Colpi:1986ye} also to model (fermion) Proca stars. Since the EOS by Colpi et al. was originally derived for a scalar field, one would then have to scale the boson mass by a factor of $1.671$ and the self-interaction by an appropriate amount. The necessary scaling for the self-interaction will be dictated by the scaling relations for pure boson stars ($M_\mathrm{max} \approx 0.22 \sqrt{\Lambda_\mathrm{int}}\, M^2_p / m$ \cite{Colpi:1986ye}) and Proca stars ($M_\mathrm{max} \approx \sqrt{\Lambda_\mathrm{int}} \ln(\Lambda_\mathrm{int})\, M^2_p / m$ \cite{Minamitsuji:2018kof}) at large self-interaction strengths. We however note that great care is needed since Proca stars technically do not exist in the limit of large self-interactions (see the analytical bound on the vector field amplitude \eqref{eq:fermion-boson-stars:fermion-proca-stars:analytical-bound-amplitude}). We plan to study this aspect in the future.

\subsection{Higher Modes and Different EOS} \label{subsec:Higher_Modes_and_different_EOS}

\begin{figure*}
\centering
    \includegraphics[width=0.495\textwidth]{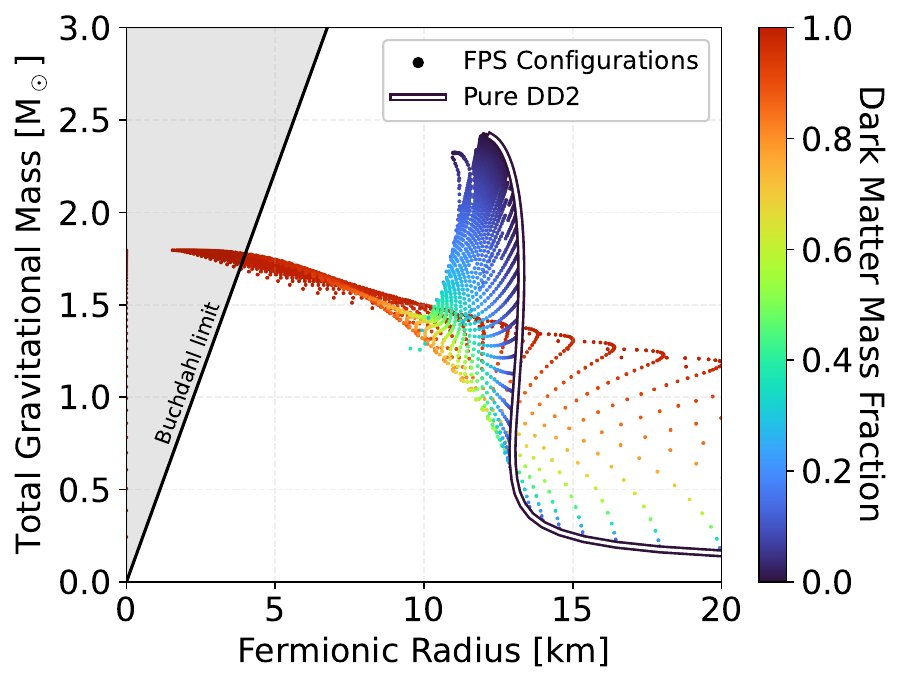}
    \includegraphics[width=0.495\textwidth]{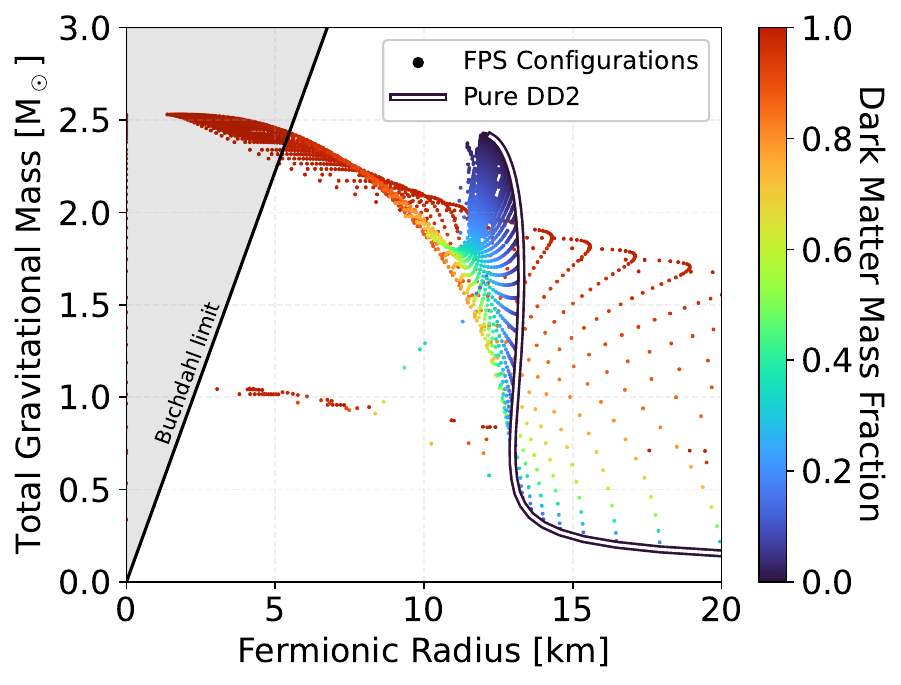}
    \caption{\textbf{Left panel:} Mass-radius diagram displaying the fermionic radius vs. the total gravitational mass for stable FPS configurations in the first mode with vector boson mass of $m = 1.34 \e{-10}\,eV$ and no self-interaction. Each point corresponds to a single configuration and is color-coded according to the DM-fraction $N_\mathrm{b}/(N_\mathrm{b} + N_\mathrm{f})$. The solid black-white line shows the mass-radius curve for pure fermionic matter.
    \textbf{Right panel:} Mass-radius diagram displaying the fermionic radius vs. the total gravitational mass for stable FPS configurations in the second mode with vector boson mass of $m = 1.34 \e{-10}\,eV$ and no self-interaction. Each point corresponds to a single configuration and is color-coded according to the DM-fraction $N_\mathrm{b}/(N_\mathrm{b} + N_\mathrm{f})$. The solid black-white line shows the mass-radius curve for pure fermionic matter.}
    \label{fig:results:fermion-proca-stars:MR-curves-mode1-mode2-comparison}
\end{figure*}


We broaden our analysis to FPSs with different EOS for the fermionic component and to FPSs where the bosonic component exists in a higher mode. Higher modes are usually assumed to be unstable, but as numerical simulations of scalar boson stars have shown \cite{DiGiovanni:2021vlu,Bernal:2009zy}, higher modes might be dynamically stable when gravitationally interacting in a multi-component system. We therefore start by considering FPSs in the first and second mode in \autoref{fig:results:fermion-proca-stars:MR-curves-mode1-mode2-comparison}. \\
In the left panel of \autoref{fig:results:fermion-proca-stars:MR-curves-mode1-mode2-comparison}, we show the total gravitational mass and the fermionic radius of stable FPS configurations, where the bosonic component is in the first mode (as opposed to the ground mode, which is the zeroth mode). The vector boson mass is $m = 1.34 \e{-10}\,eV$, and the self-interaction is set to zero. We first note the fact, that stable solutions under linear radial perturbations, according to the stability criterion \eqref{eq:fermion-boson-stars:scalar-fermion-boson-stars:FBS-stability-criterion-rewritten}, exist at all. This is a non-trivial statement as higher modes of Proca stars (and also of scalar boson stars) are usually believed to be unstable. Note however that the stability criterion \eqref{eq:fermion-boson-stars:scalar-fermion-boson-stars:FBS-stability-criterion-rewritten} is merely a necessary condition for stability and that there could be additional conditions that must be fulfilled for a solution to be stable in higher modes. Also, our stability analysis does not consider the dynamical stability of the higher modes. They might thus be unstable in non-static scenarios. It is however possible that the higher modes of the bosonic part might be stabilized through the gravitational interaction with the fermionic part of the FPS. In general, solutions in higher modes need energy to be supported or excited. Otherwise they would decay to the ground mode. The authors of \cite{Bernal:2009zy} explicitly studied the stability of higher modes for scalar boson stars with multiple scalar fields. They investigated cases where one field is in the ground mode and another is in the excited state. They found that the excited mode is stable if the charge of the conserved Noether current (which can also be interpreted as the particle number or the total restmass, see \eqref{eq:fermion-boson-stars:scalar-fermion-boson-stars:conserved-boson-number-definition}) of the ground mode is larger than the Noether charge of the excited mode: $N_{ground} > N_{excited}$. If this logic also holds for mixed systems of fermions and bosons, this would imply an additional stability condition that the fermion number $N_\mathrm{f}$ must be larger than the boson number $N_\mathrm{b}$ of the FPS in the excited mode. We further refer to chapter 3.7 in \cite{Liebling:2012fv} for a more detailed review. \\
The FPSs in the first mode exhibit higher gravitational masses in the configurations dominated by the bosonic component, compared to FPSs in the zeroth mode (compare to \autoref{fig:results:fermion-proca-stars:stability-and-MR-curve-lamda0}). The numerical value of the frequency $\omega$ in the higher mode is also larger than the frequency in lower modes. This behavior is consistent with earlier works, which studied pure Proca stars analytically \cite{Minamitsuji:2018kof} and numerically \cite{Sanchis-Gual:2017bhw}. They also observed that higher frequencies lead to larger total gravitational mass. The left panel of \autoref{fig:results:fermion-proca-stars:MR-curves-mode1-mode2-comparison} shows a number of outlier points at around $11\,km$ and $2.3\,M_\odot$. These are likely numerical artifacts due to the increased difficulty of finding accurate numerical solutions for higher modes. \\
The right panel of \autoref{fig:results:fermion-proca-stars:MR-curves-mode1-mode2-comparison} shows stable FPS configurations in the second mode. The vector boson mass is $m = 1.34 \e{-10}\,eV$  and the self-interaction is set to zero. Here also, the existence of stable solutions is to be acknowledged. In the limit of high DM-fractions, the FPSs converge to the solution of pure Proca stars and reach total gravitational masses of roughly $2.5$ times that of Proca stars in the zeroth mode (compare to \autoref{fig:results:fermion-proca-stars:stability-and-MR-curve-lamda0}). In comparison to the case in the first mode (left panel of \autoref{fig:results:fermion-proca-stars:MR-curves-mode1-mode2-comparison}), the quality of the overall solution can be seen to deteriorate further. We believe the outlier points at roughly $<13\,km$ and $1\,M_\odot$ to be non-physical numerical artifacts. The outlier points coincide with the solutions in the zeroth mode. This suggests that our solver did not find the second mode in these cases and converged on the zeroth mode instead. Solutions of FPSs in even higher modes should therefore be considered with great care. The difficulty of obtaining accurate numerical solutions is likely to increase further for higher modes. The quality of the solution is however sufficient to gain a qualitative understanding of FPSs in higher modes. In conclusion, higher modes are stable under linear radial perturbations and increase the total gravitational mass of FPSs by substantial amounts.

\begin{figure*}
\centering
    \includegraphics[width=0.499\textwidth]{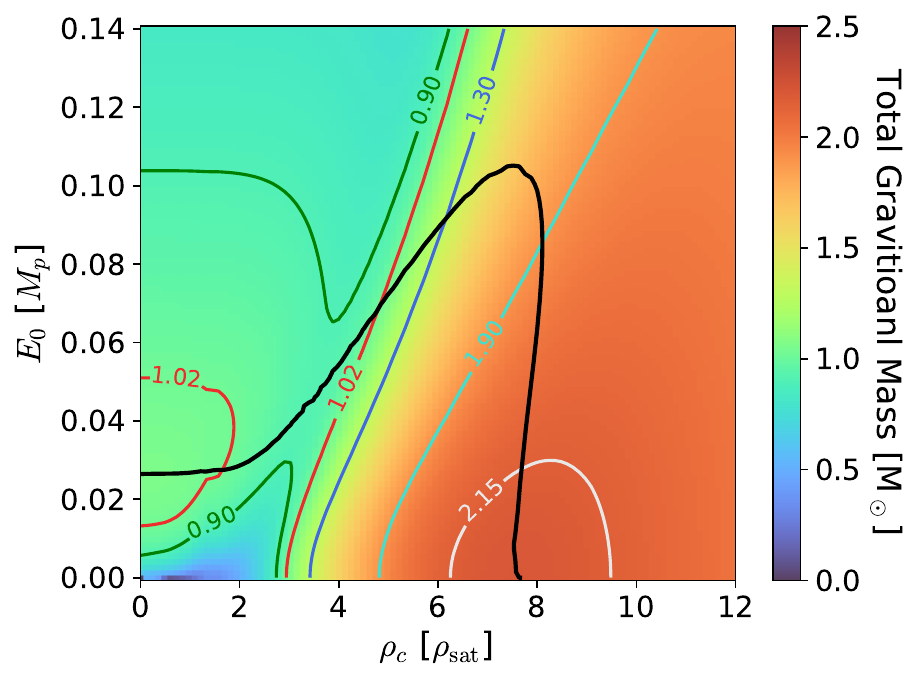}
    \includegraphics[width=0.49\textwidth]{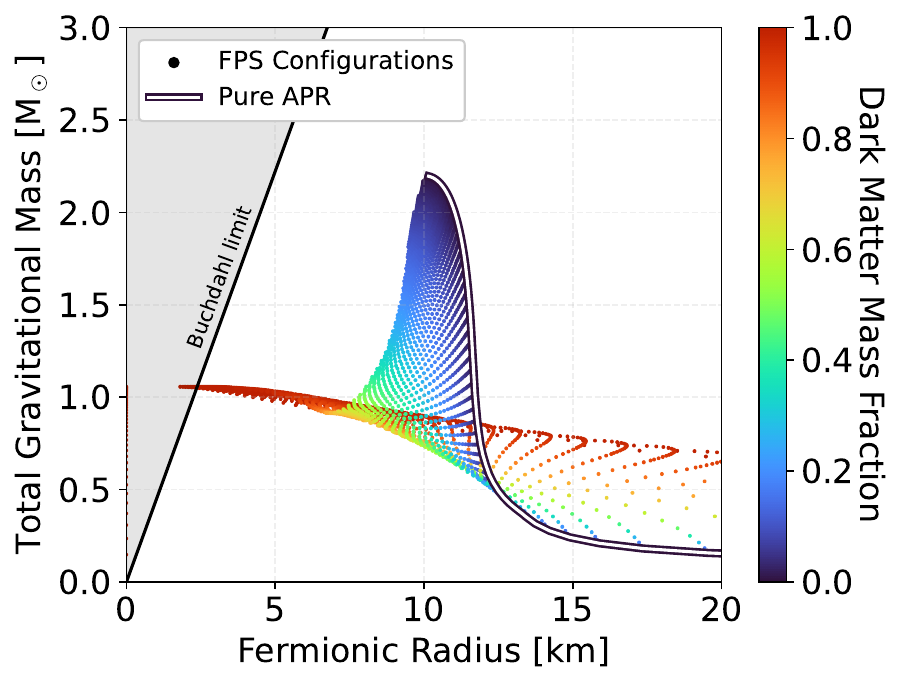}
    \caption{\textbf{Left panel:} Total gravitational mass of different FPSs as a function of the rest mass density $\rho_c$ and central vector field amplitude $E_0$. Additionally we show contours of constant gravitational mass. The black line corresponds to the stability curve, which separates stable solutions (in the lower left region) from unstable solutions (everywhere else). The qualitative behavior of the stability curve of is similar to the case with the DD2 EOS (see \autoref{fig:results:fermion-proca-stars:stability-and-MR-curve-lamda0})
    \textbf{Right panel:} Mass-radius diagram displaying the fermionic radius vs. the total gravitational mass for FPS configurations that are within the stability region displayed in the left panel. Each point corresponds to a single configuration and is color-coded according to the DM-fraction $N_\mathrm{b}/(N_\mathrm{b} + N_\mathrm{f})$. The solid black-white line shows the mass-radius curve for pure fermionic matter. In both cases, a vector field with a mass of $m=1.34 \e{-10}\,eV$ and no self-interactions was considered in addition to the APR EOS \cite{Schneider:2019vdm} for the fermionic part.}
    \label{fig:results:fermion-proca-stars:stability-and-MR-curve-APR}
\end{figure*}

\begin{figure*}
\centering
    \includegraphics[width=0.507\textwidth]{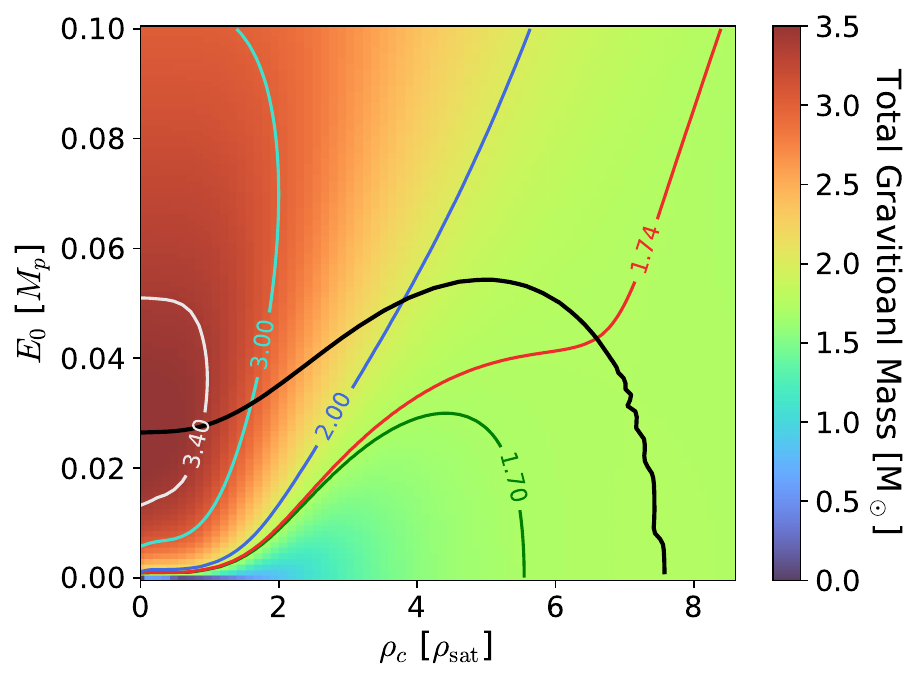}
    \includegraphics[width=0.48\textwidth]{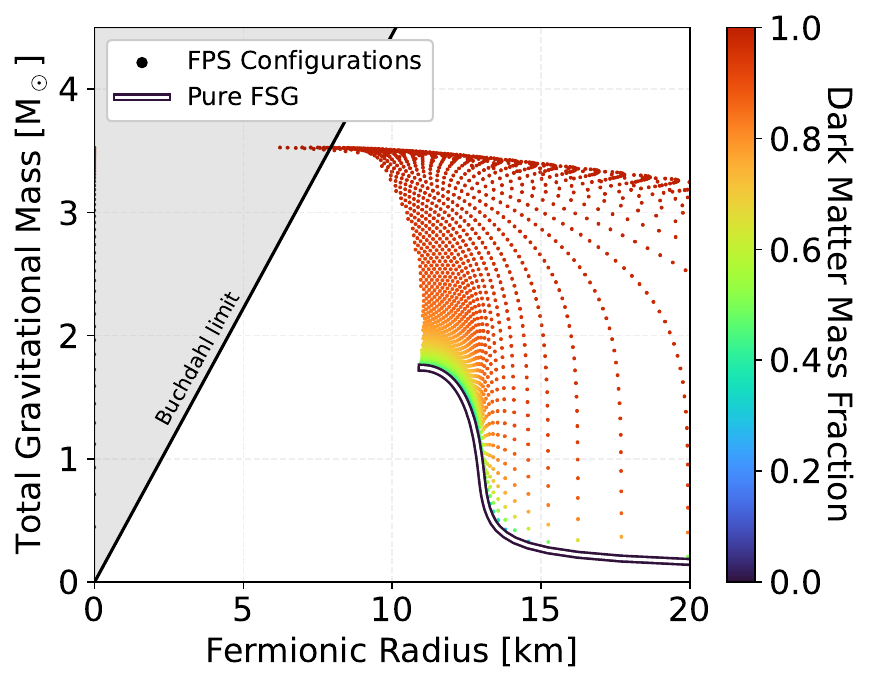}
    \caption{\textbf{Left panel:} Total gravitational mass of different FPSs as a function of the rest mass density $\rho_c$ and central vector field amplitude $E_0$. Additionally we show contours of constant gravitational mass. The black line corresponds to the stability curve, which separates stable solutions (in the lower left region) from unstable solutions (everywhere else).
    \textbf{Right panel:} Mass-radius diagram displaying the fermionic radius vs. the total gravitational mass for FPS configurations that are within the stability region displayed in the left panel. Each point corresponds to a single configuration and is color-coded according to the DM-fraction $N_\mathrm{b}/(N_\mathrm{b} + N_\mathrm{f})$. The solid black-white line shows the mass-radius curve for pure fermionic matter. In both cases, a vector field with a mass of $m=3.01 \e{-11}\,eV$ and no self-interactions was considered in addition to the FSG EOS \cite{Hempel:2009mc} for the fermionic part.}
    \label{fig:results:fermion-proca-stars:stability-and-MR-curve-FSG}
\end{figure*}

We investigate the effect that different EOS have on FPSs. In \autoref{fig:results:fermion-proca-stars:stability-and-MR-curve-APR}, we use the APR EOS \cite{Schneider:2019vdm} for the fermionic part. We chose a vector boson mass of $m=1.34 \e{-10}\,eV$ with no self-interaction for the bosonic part. In the left panel, we notice that the shape of the stability curve (black curve) is affected by the choice of the EOS. On the $\rho_c$-axis, it converges to a value of around $7.5\rho_\mathrm{sat}$. This is higher than the corresponding value of $\rho_c$ when the DD2 EOS is used (compare to \autoref{fig:results:fermion-proca-stars:stability-and-MR-curve-lamda0}) because the APR EOS is softer than the DD2 EOS. This means that the nuclear matter is easier to compress and higher central densities can be supported by the EOS. The easier compressibility also shows itself through smaller NS radii (see the right panel). In the limit of pure Proca stars, the stability curve converges to the same value as it does when the DD2 EOS is used (compare to \autoref{fig:results:fermion-proca-stars:stability-and-MR-curve-lamda0}). The MR region shows a similar qualitative behavior as in the DD2 case. The high DM-fraction limit in particular shows a convergence to the solution to pure Proca stars. The APR EOS also allows higher central amplitudes of the vector field $E_0$, compared to the DD2 EOS with equal boson mass and self-interaction strength. \\
\autoref{fig:results:fermion-proca-stars:stability-and-MR-curve-FSG} shows different FPS configurations where the FSG EOS \cite{Hempel:2009mc} was used for the fermionic part. For the bosonic part, we used a boson mass of $m=3.01 \e{-11}\,eV$ and no self-interaction. The FSG EOS is a soft EOS and thus reaches higher central densities $\rho_c$ for pure NSs. It is excluded by current observational constraints (see \autoref{fig:results:fermion-proca-stars:MR-curves-Nbfrac}), as it cannot produce pure NSs with masses of $M=2.35^{+0.17}_{-0.17}\,M_\odot$ \cite{Romani:2022jhd}. However, adding DM to the pure NSs can significantly increase the maximum gravitational mass of the combined system. The FSG EOS is then able to reach the observational bound on the maximum NS mass in the presence of DM. In fact, the MR curve of the pure DD2 EOS is entirely contained within the stability region of the FPSs with the FSG EOS. This again raises the point that some FPS solutions are degenerate with some NS solutions (see \autoref{fig:results:fermion-proca-stars:MR-curves-FBS-FPS-comparison}), when allowing for different DM-fraction and DM masses. Another factor complicating the identification of the EOS vs. the DM effects is that the presence of dark matter might change the fermionic radius produced by a given EOS. For example, \cite{Xiang:2013xwa} found that the presence of self-interacting and repulsive fermionic dark matter can lead to nearly indistinguishable fermionic radii for different EOS. To figure out whether and which types of mixed DM-NS systems might exist, it will be crucial to perform sophisticated parameter searches of the system and obtain more measurements to constrain the DM and NS properties in future studies.

%% file: conclusions.tex
In this work, we studied the impact that bosonic dark matter (DM) has on the mass and radius of neutron stars (NSs). DM was modeled as a massive, self-interacting complex vector field. DM was further assumed to only interact gravitationally with the fermionic neutron star matter. We derived the equations of motion describing static spherically symmetric fermion Proca stars (FPSs) and computed their properties numerically. We also found a scaling relation between the frequency, vector field and metric components, and we derived an analytical upper bound on the vector field amplitude. \\

We showed that the presence of the vector field can lead to core-like and to cloud-like solutions. Core-like solutions can increase the compactness of the NS component. For some configurations, observing only the fermionic radius and the total gravitational mass would appear to violate the Buchdahl limit. We found core-like solutions for vector boson masses of $m \gtrsim 1.34 \e{-10}\,eV$ and small self-interactions $\Lambda_\mathrm{int} = \lambda / 8\pi m^2$. Cloud-like solutions appeared when $m \lesssim 1.34 \e{-11}\,eV$ and $\Lambda_\mathrm{int}$ is large. For some small boson masses $m \lesssim 1.34 \e{-11}\,eV$, the presence of DM can significantly increase the total gravitational mass while leaving the fermionic radius approximately constant. \\
We computed radial profiles of FPSs and found that the existence of a maximum possible vector field amplitude limits the effect of DM on the NS when the self-interaction $\Lambda_\mathrm{int}$ is large. The maximum amplitude implies a maximum possible amount of vector boson DM accretion and could thus be used to set bounds on the DM properties. \\

We also compared FPSs to FBSs with a scalar field. We used the same parameters as in \cite{Diedrichs:2023trk} to simplify the comparison.
For stable FPS configurations, we found that many of the general qualitative trends that apply to FBSs also apply to FPSs. But vector DM leads to higher FPS masses and larger gravitational radii for equal $m$ and $\Lambda_\mathrm{int}$. This could also imply a larger tidal deformability of FPSs compared to FBSs. Also, a measurement of the gravitational radius would favor larger vector boson masses compared to scalar boson masses. \\
For FPS configurations of constant DM-fraction, we found that the effect of vector DM on the NS properties (total gravitational mass and fermionic radius) is larger compared to FBSs with equal DM-fraction, mass $m$ and self-interaction strength $\Lambda_\mathrm{int}$. One therefore needs a larger amount of scalar DM to cause the same effect as vector DM. For different boson masses and DM-fractions, we found that FPSs and FBSs can both be degenerate with each other and also be degenerate with pure NS with a different EOS. \\
We found an especially high degree of similarity between FBS solutions with no self-interaction and a boson mass of $m = 1.34 \e{-11}\,eV$ with FPS solutions where the vector boson mass is larger by a factor of $1.671$. We expect the similarity in the behavior to hold also for different boson masses (and also for non-zero self-interactions), as long as the vector boson mass is scaled accordingly by the right factor. \\
These similarities also hint towards a possibility to use the effective EOS by Colpi et al. \cite{Colpi:1986ye} also for (fermion) Proca stars. We however note that great care is needed since Proca stars do not exist in the limit of large self-interactions (see the analytical bound on the vector field amplitude \eqref{eq:fermion-boson-stars:fermion-proca-stars:analytical-bound-amplitude}). The similarities between FBSs and FPSs might also be useful for numerical applications. Scalar (fermion) boson stars are easier to implement and numerically cheaper to solve than FPSs. One could then simply solve the equations for scalar (fermion) boson stars with a re-scaled mass (and self-interaction parameter $\Lambda_\mathrm{int}$) to compute the properties ($M_\mathrm{tot}$, $R_\mathrm{f}$) of (fermion) Proca stars. \\
The prevalence of degenerate solutions highlights the importance of measuring additional observables, such as the tidal deformability, to break the degeneracies. \\

We confirmed the existence of higher modes that are stable under first-order radial perturbations. We found that higher modes lead to higher total gravitational masses of the mixed FPS systems. Using FPSs with different EOS for the fermionic part, we explicitly confirmed that for certain DM masses, previously excluded EOS are able to fulfill observational bounds if DM is present. Mixed systems of bosonic DM and NS matter can therefore be consistent with all current observational constraints if suitable boson masses and self-interaction strengths are chosen.

%% file: appendix.tex
\section{Units} \label{sec:appendix:units}

In this work, we use units in which the gravitational constant, the speed of light and the solar mass are set to $G = c = M_\odot = 1$. As a direct consequence, distances are measured in units of $\approx 1.48\,km$, which corresponds to half the Schwarzschild radius of the Sun (also called the gravitational radius of the Sun). The Planck mass is $M_p = \sqrt{\hbar c / G} \approx 1.1 \times 10^{-38} M_\odot$. Since $G = c = M_\odot = 1$ it follows that $\hbar \approx 1.2 \times 10^{-76} \neq 1$. \\
Boson stars (with a scalar field) are described using the Klein-Gordon equation, which in SI units and flat spacetime reads $(\square - (mc/\hbar)^2)\phi = 0$. The term $mc/ \hbar$ is the inverse of the reduced Compton wavelength $\lambda_c = \hbar /mc$, which sets the typical length scale for the system even in the self-gravitating case. We assume that the typical length scale of the boson is similar to the gravitational radius $GM_\odot/c^2$, which in the case of mass scales of $\sim 1\,M_\odot$ is approximately $1.48\,km$. With $m = \hbar / c \lambda_c$, this therefore leads to a mass scale of the bosonic particle of $1.336 \e{-10}\,eV$. Previous works such as, e.g., \cite{Diedrichs:2023trk,DiGiovanni:2021ejn} thus specify the mass of the scalar particle in these units. A mass of $m=1$ in our numerical code \cite{Diedrichs-Becker-Jockel} then also corresponds to $1.336 \e{-10}\,eV$. This choice of the boson mass then automatically leads to boson stars with masses in the range of $\sim 1\,M_\odot$. The same reasoning can also be applied to the case where the boson is a vector boson. This is valid since all components of a vector field also fulfill the Klein-Gordon equations individually.